\documentclass[english,useAMS, usenatbib]{mn2e}
\usepackage[T1]{fontenc}
\usepackage[latin9]{inputenc}
\usepackage{units}
\usepackage{rotating}
\usepackage{url}
\usepackage{amsmath}
\usepackage{amssymb}
\usepackage{graphicx}
\usepackage{esint}
\usepackage[authoryear]{natbib}
\usepackage{listings}

\makeatletter

\def\gtsima{$\; \buildrel > \over \sim \;$}
\def\ltsima{$\; \buildrel < \over \sim \;$}
\def\prosima{$\; \buildrel \propto \over \sim \;$}
\def\gsim{\lower.5ex\hbox{\gtsima}}
\def\lsim{\lower.5ex\hbox{\ltsima}}
\def\simgt{\lower.5ex\hbox{\gtsima}}
\def\simlt{\lower.5ex\hbox{\ltsima}}
\def\simpr{\lower.5ex\hbox{\prosima}}

\providecommand{\tabularnewline}{\\}

\title[CRASH with metals]{CRASH3: cosmological radiative transfer through metals}
\author[L. Graziani et al.]{L. Graziani$^{1}$
\thanks{E-mail: lgmaxp@mpa-garching.mpg.de }, A. Maselli$^{2}$, B. Ciardi$^{1}$\\
$^{1}$Max-Planck-Institut f\"ur Astrophysik, Karl-Schwarzschild-Stra{\ss}e 1, D-85748 Garching b. M\"unchen, Germany \\
$^{2}$EVENT Lab for Neuroscience and Technology, Universitat de Barcelona, Passeig de la Vall d'Hebron 171, 08035 Barcelona, Spain}

\usepackage{babel}

\begin{document}

\date{Accepted 2010 <Month> XX. Received 2010 <Month> XX; in original form
2010 <Month> XX}

\maketitle
\pagerange{\pageref{firstpage}--\pageref{lastpage}} \pubyear{2010}\label{dat:firstpage}

\begin{abstract}
Here we introduce \texttt{CRASH3}, the latest release of the 3D radiative transfer code \texttt{CRASH}.
In its current implementation \texttt{CRASH3} integrates into the reference algorithm the code \texttt{Cloudy} to evaluate the ionisation states of metals, self-consistently with the radiative transfer through H and He.  The feedback of the heavy elements on the calculation of the gas temperature is also taken into account, making of {\tt CRASH3} the first 3D code for cosmological applications which treats self-consistently the  radiative transfer through an inhomogeneous distribution of metal enriched gas with an arbitrary number of point sources and/or a background radiation.
The code has been tested in idealized configurations, as well as in a more realistic case of multiple sources
embedded in a polluted cosmic web.
Through these validation tests the new method has been proven to be numerically stable and 
convergent.
We have studied the dependence of the results on a number of physical quantities such as  the source characteristics (spectral range and shape, intensity), the metal composition, the gas number density and metallicity.

\end{abstract}


\section{INTRODUCTION\label{sec:INTRODUCTION}}

During the last decade, observational and theoretical studies constraining
the nature of the intergalactic medium (IGM) have shown that metals
are a pervasive component of the baryonic budget of our universe and
that they are associated with a wide range of hydrogen column density
($N_{\textrm{HI}}$) systems typically identified in quasars (QSO)
absorption spectra at different redshift (\citealt{b46-Meyer1987,b47-Lu1991,b44-Songaila1996,b45-Cowie1998,b48-Ellison2000,b50-Songaila2001,b57-Schaye2000,b49-Schaye2003,b56-Aracil2004,b58-Pieri2004,b4-Becker2011};
also see \citealt{b27-Meikson2009} for a review).

High H$\, \rm \scriptstyle I$ column density absorbers, with $N_{\textrm{HI}}\geqslant2\cdot10^{20}\textrm{cm}^{-2}$,
are easily identified in these spectra for the presence of strong
damping wings, and are classified as Damped Ly$\alpha$ systems (DLAs). Column
densities in the range $1.6\cdot10^{17}\,\textrm{cm}^{-2}\leq N_{\textrm{HI}}\leq10^{20}\,\textrm{cm}^{-2}$
are instead associated with Lyman Limit Systems (LLSs) and spectroscopically
identified by their strongly saturated Ly$\alpha$ lines. Both systems
show C$\, \rm \scriptstyle IV$ lines as well as many ions in lower ionisation states (Mn$\, \rm \scriptstyle II$,
Si$\, \rm \scriptstyle II$, Fe$\, \rm \scriptstyle II$), and
are typically associated with metallicities of $Z \sim 10^{-2}Z_{\odot}$ (LLSs;
e.g. \citealt{b73-Steidel1990}) and $10^{-2}Z_{\odot}\leq Z\leq0.3Z_{\odot}$
(DLAs; e.g. \citealt{b73-Steidel1990,b16-Hellsten1997,b53-Rauch1997,b75-Pettini1999,b74-Kulkarni2005}). 

The presence of metals in LLSs and DLAs can be interpreted as a natural
product of the stellar nucleosynthesis acting therein. LLSs are in
fact identified as clouds in the galactic halos, while high redshift
($z\sim3$) DLAs are believed to be the progenitors of the present-day
galaxies. 

Advances in high resolution spectroscopy revealed 
weak metal absorption lines in the
Ly$\alpha$ forest ($N_{\rm HI} \simlt 10^{-17}$~cm$^{-2}$), with C$\, \rm \scriptstyle IV$
detected in most of the systems with $N_{\textrm{HI}}\geqslant10^{15}\textrm{cm}^{-2}$
and in more than half of the systems with $N_{\textrm{HI}}\geqslant10^{14}\textrm{cm}^{-2}$
(\citealt{b51-Tytler1995,b44-Songaila1996}). The typical metallicities of the Ly$\alpha$ forest are estimated
in the range $10^{-4}Z_{\odot}\leq Z\leq10^{-2}Z_{\odot}$ (\citealt{b59-Simcoe2004}).
The subsequent discovery of a metallic component in less dense regions
(\citealt{b45-Cowie1998,b48-Ellison2000,b57-Schaye2000,b49-Schaye2003,b56-Aracil2004,b58-Pieri2004}) has been interpreted as
the evidence of efficient spreading mechanisms which are able to transport the metals far away
from their production sites and pollute low density regions.

The redshift evolution of the gas metallicity has been extensively
investigated. In the redshift range $1.5<z<4$, C$\, \rm \scriptstyle IV$
and Si$\, \rm \scriptstyle IV$ doublets are the main tracers of
the IGM metallicity, because their rest frame wavelength is larger 
than the $\textrm{Ly}\alpha$ and the lines cannot be confused with those of
the forest. In this range the column density distribution of C$\, \rm \scriptstyle IV$
seems to remain constant (\citealt{b50-Songaila2001}). 
Although a consensus has not been reached yet, a decline in the abundance of
C$\, \rm \scriptstyle IV$ above $z\sim4.5$ is reported by different
groups (\citealt{b33-Ryan-Weber2009,b3-Becker2009,b4-Becker2011,b93-Simcoe2011}).
The Si$\, \rm \scriptstyle IV$ doublet $\lambda\lambda1393.76,1402.77$ \AA\ has been investigated in QSO line  surveys in the redshift range 1.5 < z < 5.5. 
\cite{b50-Songaila2001} measured $\Omega_{\rm Si\, \rm \scriptstyle IV}$, the Si$\, \rm \scriptstyle IV$ 
mass density relative to the critical density, in the redshift range 
$2 < z < 5.5$, finding a roughly constant value in the range  $2<z<4.5$ for absorbers with column 
densities $10^{13} {\rm cm}^{-2} \leq N_{\rm Si\, \rm \scriptstyle IV} \leq 10^{15} {\rm cm}^{-2}$, 
while $\Omega_{\rm Si\, \rm \scriptstyle IV}$ may have increased by an order of magnitude in $4.5 <z< 5.5$. 
These abundance trends have been confirmed by subsequent studies (\citealt{b95-Boksenberg2003,b36-Songaila2005,b61-Scannapieco2006}).

At higher redshifts, a critical diagnostic
role could be played instead by ions with lower ionisation states like 
C$\, \rm \scriptstyle II$,
Si$\, \rm \scriptstyle II$ and O$\, \rm \scriptstyle II$, probing colder gas (\citealt{b31-Oh2002,b15-Furlanetto2003,b4-Becker2011}).

Below $z\sim1$ a larger sample of data is available also for the oxygen component which is the major tracer of the IGM metallicity. The interested reader can find more information in the recent study of \citet{b94-Cooksey2011}, and references therein.

Beyond this general consensus, many observational data remain controversial,
as well as their interpretation which is of primary importance, e.g.
in constraining different enrichment scenarios. The metal abundance,
the number of ionisation states and the distribution in space
and time are still subjects of intense debate (see \citealt{b55-Petitjean2001,b54-Petitjean2005}).

The theoretical interpretation of the data is of particular relevance
because observations of metals in dense regions, where stellar nucleosynthesis
is active, provide a record of the star formation history, while observations
in the IGM can be a good indicator of the galactic winds efficiency,
the velocity structure of the IGM and more in general of the enrichment
mechanism (\citealt{b62-Gnedin1997,b64-Ferrara2000,b63-Cen2001,b20-Madau2001}). In fact, different scenarios
have been considered in the literature, as an early enrichment by the
first generation of stars \citep{b20-Madau2001}, a continuous enrichment 
(\citealt{b62-Gnedin1997,b64-Ferrara2000,b65-Scannapieco2002}),
or a late enrichment coinciding with the star formation peak at $z\sim2-4$
\citep{b1-Adelberger2002}. 
In addition, the determination of metal abundances could also constrain
the efficiency of the gas cooling function \citep{b37-Sutherland1993,b81-Maio2007,b80-Smith2008,b79-Wiersma2009a}
and the formation of massive galaxies \citep{b67-Thacker2002}.

For these reasons, several theoretical schemes for metal production and spreading
have been developed, based on both semi-analytic
models (e.g. \citealt{b65-Scannapieco2002, b61-Scannapieco2006}) and numerical simulations (\citealt{b46a-Mosconi2001,b65a-Scannapieco2005, b28-Oppenheimer2006,b92-Dubois2007,b76-Wiersma2009b,b82-Maio2010,b85-Schaye2010,b34-Shen2010,b39-Tornatore2010,b78-Wiersma2010,b83-Maio2011,b77-Wiersma2011}).
In the most advanced approaches the ionisation state of the metals is calculated
assuming the presence of a uniform UV background, which is then used
as energetic input for photo-ionisation codes computing the metal
ionisation states at the equilibrium (\citealt{b28-Oppenheimer2006,b29-Oppenheimer2009,b96-Oppenheimer2012}). The results of this approach
are affected by the uncertainties associated to the assumptions on
the shape and intensity of the radiation field, which is not calculated
self-consistently from the radiative transfer across the inhomogeneous
gas distribution. Many studies suggest in fact that shadowing, filtering
and self-shielding induce deviations in the shape and intensity of
the background with respect to models in which the effects of the
radiative transfer are neglected (\citealt{b25-MAselli2005} and references therein).

Fluctuations in the photo-ionisation rates as well as spatial deviations
in the IGM temperature due to the inhomogeneity of the cosmic web
support this view at least on scales of few co-moving Mpc (see \citealt{b25-MAselli2005,b14-Furlanetto2009,b27-Meikson2009}
and references therein). On larger scales ($\sim100h^{-1}$ Mpc co-moving)
the source spatial distribution and their spectral variability could
be an additional cause of variations in the UVB                            
(\citealt{b90-Zuo1992b,b91-Zuo1992a,b89-Meiksin2003,b4-Bolton2011}). 

The fluctuations induced by radiative transfer effects could also
be efficiently recorded in the ionisation state of the metals, because
their rich electronic structure and atomic spectrum are more sensitive,
compared to H and He,
to the radiation field fluctuations (\citealt{b31-Oh2002,b15-Furlanetto2003,b14-Furlanetto2009}).

Numerical schemes which solve the cosmological radiative transfer
equation by applying different approximations are now quite mature
and well tested (\citealt{b18-Iliev2006a,b18-Iliev2009} and references therein for an
overview of the available codes) and are able to simulate complex
scenarios involving large cosmological boxes and number of sources
 (e.g. \citealt{b7-Ciardi2003a,b8-Ciardi2003b,b68-Iliev2006b,b70-Trac2007,b26-McQuinn2009,b87-Baek2010,b84-Ciardi2011}). Typically, these codes
are restricted to the hydrogen chemistry, with only a few of them
including a self-consistent treatment of the helium component, which
is particularly relevant for a correct determination of the gas temperature
(see e.g. \citealt{b84-Ciardi2011}). None of them though includes the treatment
of metal species. 

In the interstellar medium (ISM) community, on the other hand, several photo-ionisation
codes, as \texttt{Cloudy} \citep{b12-Ferland1998}, \texttt{MAPPINGSIII} \citep{b86-Allen1998} 
and \texttt{MOCASSIN} \citep{b300-Ercolano2003} are able to simulate the complex 
physics of galactic HII regions largely polluted by heavy elements.

The aim of the present work is to describe a novel extension of the
radiative transfer code\texttt{ CRASH} \citep{b6-Ciardi2001,b24-Maselli2003,b23-Maselli2009}, which has been integrated
with \texttt{Cloudy}, allowing the prediction of metal ionisation
states self-consistently with the radiation field as calculated
by the radiative transfer through the IGM density
field. 

The paper is structured as follows. In Section 2 and 3 we briefly
introduce the radiative transfer code \texttt{CRASH} and the photo-ionisation
code \texttt{Cloudy}. In Section 4 we illustrate the details of the
integration of the two codes into a self-consistent pipeline. The 
tests of the new \texttt{CRASH} variant called \texttt{CRASH3} are
reported in Section 5. Section 6 summarizes the conclusions.

\section{CRASH\label{sec:CRASH2}}

\texttt{CRASH} is a 3D radiative transfer (RT) code designed to follow
the propagation of hydrogen ionising photons (i.e. with energy $E\geq13.6$~eV)
through a gas composed by H and He. The code adopts a combination
of ray tracing and Monte Carlo (MC) sampling techniques to propagate
photon packets through an arbitrary gas distribution mapped on a cartesian
grid, and to follow in each grid cell the evolution of gas ionisation
and temperature. This treatment guarantees a reliable description
of such evolution in a large variety of configurations, as shown by
the Cosmological Radiative Transfer Comparison Project tests \citep{b18-Iliev2006a}
and its various applications to the study of the H and He reionisation
(\citealt{b7-Ciardi2003a,b8-Ciardi2003b, b84-Ciardi2011}), the imprints of the fluctuating background
on the $\textrm{Ly}\alpha$ forest \citep{b25-MAselli2005}, the quasar proximity
effects \citep{b92-Maselli2007,b92-Maselli2009} and the impact of reionisation on the visibility 
of $\textrm{Ly}\alpha$ emitters (\citealt{b100-Dayal2011,b101-Jeeson-Daniel2012}).

The MC algorithm adopted allows to easily add new physical processes.
In its first version \citep{b6-Ciardi2001} the code describes H photo-ionisation
due to point sources, and includes the effect of re-emission following
gas recombination. The subsequent versions brought about significant
improvements. First the physics of He and the thermal evolution of
the gas have been introduced, together with the treatment of an ionising
background field \citep{b24-Maselli2003,b25-MAselli2005}. In the latest release ({\tt CRASH2}; \citealt{b23-Maselli2009},
hereafter MCK09) we introduced multi-frequency photon packets obtaining significant
improvements in terms of accuracy of the ionisation and temperature
profiles, as well as computational speed. Hereafter the code name
\texttt{CRASH} will refer to the version \texttt{CRASH2}. 

In parallel with the reference code, variants and extensions have
been developed such as: MCLy$\alpha$ \citep{b40-Verhamme2006}, which has adapted
the reference algorithm to treat the resonant propagation of $\textrm{Ly}\alpha$
photons; \texttt{CRASH}$\alpha$ (\citealt{b41-Pierleoni2009}), which follows
the self-consistent propagation of both $\textrm{Ly}\alpha$ photons and ionising
continuum radiation; \citet{b72-Partl2011} have instead developed an MPI parallel
implementation of the base \texttt{CRASH} algorithm.

The new release described in this paper (hereafter \texttt{CRASH3})
extends the standard \texttt{CRASH} photo-ionisation algorithm to
the treatment of the most cosmologically relevant metals in atomic form: C, O and
Si. The current numerical scheme is based on a new pipeline which
combines the excellent capabilities of \texttt{CRASH} in tracing the
radiation field with the sophisticated features of the photo-ionisation
software \texttt{Cloudy} \citep{b12-Ferland1998}. The inclusion of a considerably
large set of data imposed by the numerous metal ionisation states 
has required a substantial source code re-engineering that introduces
a new memory management and a re-modelling of the photon packet-to-cell
interaction. \texttt{CRASH3} is consequently more modular, optimized
and easily integrable with other codes. In the following, we           
briefly review the basic ingredients of the \texttt{CRASH} algorithm
that are required to understand the implementation of the new version,
and we refer the reader to the original papers for more specific details. 

A \texttt{CRASH} simulation is defined by assigning the initial conditions
(ICs) on a regular three-dimensional cartesian grid of a given dimension ($N_{c}^{3}$
cells) and a physical box linear size of $L_{b}$, specifying: 
\begin{itemize}
\item the number density of H ($n_{{\rm H}}$) and He ($n_{{\rm He}}$), the gas temperature
($T$) and the ionisation fractions ($x_{\rm HII}=n_{\rm HII}/n_{\rm H}$,
$x_{\rm HeII}=n_{\rm HeII}/n_{\rm He}$ and $x_{\rm HeIII}=n_{\rm HeIII}/n_{\rm He}$)
at the initial time $t_{0}$; 
\item the number of ionising point sources ($N_{s}$), their position in
cartesian coordinates, luminosity ($L_{s}$ in $\textrm{erg}\,\textrm{s}^{-1}$)
and spectral energy distribution (SED, $S_{s}$ in erg~s$^{-1}$~Hz$^{-1}$)  
assigned as an array whose element provide the relative intensity of the radiation 
in the correspondent frequency bin (see MCK09 for more details);
\item the simulation duration $t_{f}$ and a given set of intermediate times
$t_{j}\in\left\{ t_{0},\ldots,t_{f}\right\} $ to store the values
of the relevant physical quantities;
\item the intensity and spectral energy distribution of a background radiation,
if present. 
\end{itemize}
The simulation run consists in emitting photon packets from the ionising
sources and following their propagation through the domain. Each photon
packet keeps traveling and depositing ionising photons in the crossed
cells, as far as its content in photons is completely extinguished
or it escapes from the simulated box (although periodic boundary conditions
in the packets propagation are possible).

At each cell crossing, \texttt{CRASH} simulates the radiation-to-gas
interaction by evaluating the absorption probability for a single
photon packet as:
\begin{equation}
P\left(\tau\right)=1-e^{-\tau},
\label{eq: OPTTHIN}
\end{equation}
where $\tau$ is the total gas optical depth of the cell given by the
sum of the contributions from the different species, i.e. $\tau=\tau_{\textrm{HI}}+\tau_{\textrm{HeI}}+\tau_{\textrm{HeII}}$.
The number of photons absorbed in the cell can then be estimated as:
\begin{equation}
N_{\gamma}\left(1-e^{-\tau}\right),
\label{eq:ABSPHOTONSCELL}
\end{equation}
where $N_{\gamma}$ indicates the photon content of a packet entering
the cell.

The number of the deposited photons for each spectral frequency is
then used to compute the contribution of photo-ionisation and photo-heating
to the evolution of $x_{\textrm{HII}}$, $x_{\textrm{HeII}}$,
$x_{\textrm{HeIII}}$ and $T$. The set of equations
solved in the code is described in detail in \citet{b24-Maselli2003} and in
MCK09.

For the sake of the following discussion, we remind here that the
evolution of the thermal state of the gas in each cell is regulated
by the formula:
\begin{equation}
\frac{dT}{dt}=\frac{2}{3k_{B}p}\left[k_{B}T\frac{dp}{dt}+\mathbb{{\cal H}}\left(T,x_i\right)-\Lambda\left(T,x_i\right)\right],
\label{eq:ENBALANCECELL}
\end{equation}
where $p$ is the number of free particles per unit volume, ${\cal H}$
and $\Lambda$ are respectively the heating and the cooling functions
and the subscript $i$ refers to all the ion species composing the gas.
$k_{B}$ is the Boltzmann constant. The heating function is determined
by computing the photo-heating resulting from the photon-to-gas interaction discussed above,
while $\Lambda$ is calculated by adding up the contribution of
various radiative processes: collisional ionisation and excitation,
recombinations, Bremsstrahlung and Compton cooling. Differently from
photo-heating, these processes are treated as continuous processes,
described by their respective rates (see \citealt{b24-Maselli2003} for details).

\section{CLOUDY \label{sec:Cloudy}}

\texttt{Cloudy} \citep{b12-Ferland1998} is a code designed to simulate the physics
of the photo-ionised regions produced by a wide class of sources ranging
from the high temperature blue stars to the strong X-ray emitting
Active Galactic Nuclei. The main goal of \texttt{Cloudy} is the prediction
of the physical state of photo-ionised clouds including all the observably
accessible spectral lines. The latest stable release of \texttt{Cloudy}
(at the time of writing v. 10.0%
\footnote{\url{http://www.nublado.org}%
}) simulates a gas which includes all the heavy elements of the typical
solar composition and the contribution of dust grains and molecules
present in the ISM.

In this Section we will focus on the description of the \texttt{Cloudy}
features that have been primarily used to implement \texttt{CRASH3}.
The reader interested in the details of the code implementation or
in reviewing the many physical processes included can
find more appropriate references in \citet{b12-Ferland1998} and \citet{b11-Ferland2003}.

Unlike \texttt{CRASH}, \texttt{Cloudy} is a 1D code assuming as preferred
geometrical configuration a symmetrical gas distribution around a
single emitting source, with photons propagating along the radial
direction. \texttt{Cloudy} can also simulate the diffuse continuum
re-emitted by recombining gas as nearly isotropic component under the
assumption that the diffuse field contribution is generally small
and can be treated by lower order approximations. Additional isotropic
background fields can also be handled, as long as their shape and
intensity are specified by the user. Some popular background models
(like the Haardt and Madau cosmic UV spectrum described in \citealt{b22-Madau2009}) are already
distributed with the code. The contribution of the Cosmic Microwave Background
(CMB) radiation can also be
accounted for because it is an important source
of Compton cooling for low density gas configurations typical of the
IGM.

The micro-physics implemented in \texttt{Cloudy} is very accurate:
it includes all the metals present in the typical solar composition
\citep{b71-Grevesse1998} described as multi-level systems and treated self-consistently
with the ions of the lightest 30 elements. Photo-ionisation from valence,
inner shells and many excited states, as well as collisional ionisation
by both thermal and supra-thermal electrons and charge transfer, are
included as ionisation mechanisms. The gas recombination physics is
simulated including the charge exchange, radiative recombination,
and dielectronic recombination processes. \texttt{Cloudy} simulates
all these processes adopting an approximation method for the radiation
field evaluation known as escape probabilities method \citep{b17-Hubeny2001},
instead of evaluating the full radiative transfer as done by \texttt{CRASH}.
This choice implies the loss of many details pertaining the line properties
description, e.g. line pumping by the incident continuum, photon destruction
by collisional deactivation and line overlap. In the standard release
of \texttt{CRASH} the details of the lines are not accounted for and
the above limitations are thus negligible. 

Once the characteristics of the source and the species involved in
the calculation are set up, \texttt{Cloudy} estimates the radial distribution
of the ionisation and temperature fields by simultaneously solving
the equations of ionisation and thermal equilibrium (\citealt{b10-Dopita2003,b30-Osterbrock2005}).
A static solution describing the physical state of the gas is then
found by dividing the domain in smaller regions and iterating until
convergence is reached. The usual assumption of such calculations
is that atomic processes occur on timescales much shorter than the
temporal scales of the macro-physical system. \texttt{Cloudy} does
not 'a priori' assume that the gas is in equilibrium and the solution
provided is a general non-equilibrium ionisation configuration for
a static medium that does not retain any information of the temporal
evolution of the system towards the converging state. A large set
of information about the relevance of the physical processes that
establish the final convergence, and the details of the line emission
processes are also provided with a great level of detail. This micro-physic
treatment can not be directly handled in \texttt{CRASH }with the same
level of accuracy.

\texttt{Cloudy} describes the thermal equilibrium of the photo-ionised
gas providing the local thermal balance obtained in each simulated
sub-region. In the absence of non-thermal electrons produced by high-energy
photons, this thermodynamic equilibrium is generally specified by
the electron temperature $T_{e}$, because the electron velocity distribution
of the gas is predominantly Maxwellian. 
Despite {\tt CRASH} does not distinguish between the contribution of the
different species to the gas temperature as calculated in eq.~\ref{eq:ENBALANCECELL},
in practice $T \sim T_e$ to a good approximation, and thus it is 
consistent with the output from {\tt Cloudy}. It should be clarified though
that {\tt Cloudy} provides a gradient of temperatures within the simulated region. Because the {\tt CRASH} resolution is a single cell, whenever
we use a temperature provided by {\tt Cloudy} this should be intended as
the volume average over a cell. In the following we will always refer to the
temperature as $T$.

Implementing {\tt CRASH3} we have minimised the differences between the two codes,
e.g. by disabling in {\tt Cloudy} all the physical processes non
explicitly treated in {\tt CRASH} such as molecule and dust grain
physics, charge transfer effects and radiation pressure, among others. 
In addition, while \texttt{CRASH} simulates the propagation of
hydrogen ionising photons, \texttt{Cloudy} requires that any spectral
information is provided in the energy range $13.6136\cdot10^{-8}$~eV$<E<100$~MeV.
For this reason, the spectrum used as input for \texttt{Cloudy} is
the one provided by {\tt CRASH} in the frequency range
13.6~eV$\leq E\leq E_{max}$ (see following Section), while it is set to zero for
$13.6136\cdot10^{-8}$~eV$<E<$13.6~eV and $E_{max}<E<$100~MeV, where $E_{max}$ is the energy corresponding to the higher frequency sampled in the \texttt{CRASH} simulation (see MCK09 for more details).
The deeply different set-up of the simulations in \texttt{CRASH} and
\texttt{Cloudy}, both in geometrical configuration and in the
gas micro-physics, has made the building of the common pipeline a non
trivial task. This will be detailed in the following Sections.

\section{THE INCLUSION OF METALS IN CRASH\label{sec:METALS-IN-CRASH}}

The extension of the \texttt{CRASH} algorithm with the full micro-physics
of metals is hardly viable because of the extreme complexity of the
metal electronic structure which would increase the computational
costs of a RT simulation to unacceptably high levels (see \citealt{b10-Dopita2003,b88-Draine2011}
and references therein for a modern introduction to the physics of
the metals in ionised regions). 

\begin{figure}
\centering
\includegraphics[scale=0.75]{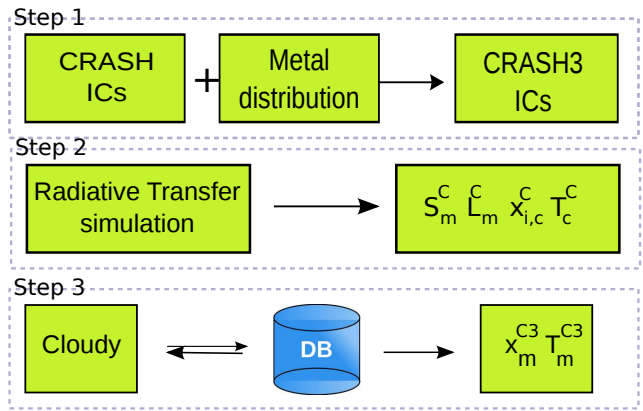}
\caption{\texttt{CRASH3} simulation pipeline.
  The quantities $S$, $L$, $x_i$ and $T$ are the SED,
  luminosity, ionisation fractions (per species $i$) and gas temperature. The subscripts $c$ and $m$ refer
  to all the cells in the computational domain and the metal enriched cells only, while
  the superscripts $^C$ and $^{C3}$ refer respectively to the quantities provided
  by the radiative transfer simulation in Step 2 and by the database (DB), i.e. as a
  product of the full {\tt CRASH3} pipeline.
  See text for more details.}
\label{fig:CRASH-PIPELINE}
\end{figure}

For this reason, we have adopted a hybrid approach in which \texttt{CRASH}
performs the RT only through H and He, while the metal ionisation
states are implemented self-consistently, but they are calculated
with \texttt{Cloudy}. The two algorithms
interact within a single pipeline called \texttt{CRASH3}, which is
sketched in Figure~\ref{fig:CRASH-PIPELINE}.
It should be noted that, if the
radiative transfer is performed in a cosmological context, the pipeline
applies to each single redshift, $z$. In this case, in addition to
the ICs of the RT simulations, other physical quantities might depend
on $z$, e.g. the cooling off the CMB radiation.

More specifically, \texttt{CRASH3} recognises
the metal enriched sub-domain by applying a masking technique to it
and by labelling the enriched cells, which are indicated here with the subscript $m$. 
After masking the $m$-cells containing metals,
it derives the spectral energy distribution and
luminosity of the ionising radiation present in each of the $m$-cells
($S,L$)$_m$, which are then used to query a dynamic database (DB)
of configurations pre-computed with {\tt Cloudy} to identify the corresponding 
ionisation states of the metals and the temperature of the gas. 
This procedure is repeated for a set of times 
$t_{k}\in\left\{ t_{0},\ldots,t_{f}\right\} $
and the physical state the enriched cells is then evaluated as a 
sequence of $k_f$ equilibrium configurations at $t_k$.
To ensure that the ionisation
level of the metals is consistent with the energy configuration,
the pipeline checks that the ionisation fractions of H and He evaluated
by \texttt{Cloudy} and \texttt{CRASH} are in agreement.
An example of this internal convergence test can be found
in Section~\ref{sub:MetalCRASH-vs-CRASH2}.

It should be noted that this approach neglects the contribution of
metals to the optical depth, as well as to the diffuse radiation 
emitted by recombining gas. This
approximation is justified as long as the metal abundances are very
small compared to those of H and He. In fact, when the number
density of heavy elements $n_{\textrm{Z}} \simgt 10{}^{-3}n_{\textrm{H}}$,
the total photo-ionisation cross section starts to be dominated by the metal
component (see \citealt{b88-Draine2011}, Chapter thirteen for a complete discussion).
Because the metallicity observed in the IGM is typically below this value (see
the Introduction), the above assumption is justified in the
cases of our interest. 

As detailed in Section~3, \texttt{Cloudy} can deal with either a
single source configuration or a background radiation. If only a single
source is present in the computational domain of \texttt{CRASH} or
when each cell is illuminated from a preferential direction, then
the \texttt{Cloudy} point source configuration should be used 
to estimate the ionisation and temperature
state of the cell. When instead a cell is illuminated more
or less uniformly from all directions, then the background radiation
set-up should be adopted.

In the current implementation of \texttt{CRASH3} we have included
the species C, O and Si, which are the most abundant metals observed
in the IGM (see the Introduction). However, the inclusion of other
species, which may be relevant for a more accurate estimate of the
gas cooling function, is a straightforward extension of the present
scheme (see Sec.~5.1.5).

Despite the conceptual simplicity of the approach described above, the coupling of
\texttt{CRASH} and \texttt{Cloudy} in a single numerical scheme presents several technical challenges. In the following we will describe in more detail
the strategy adopted for such integration and some aspects of the pipeline.

\subsection{Step 1: initial conditions}

The starting point of the pipeline is the set-up of the \texttt{CRASH3} ICs, 
which are the same of \texttt{CRASH} as specified in Section~2, with the 
addition of the spatial distribution and abundance of
all the metal species present in the computational domain. These can
be artificially created by hand or can be
obtained as a result of e.g. hydrodynamic simulations that include
physical prescriptions for metal production and spreading.

A preliminary analysis of the spatial distribution of metals allows
to identify the $m$-cells that need to track the radiation field 
(Step 2) and then query the pre-computed DB (Step 3). To
identify these cells a boolean mask is built isolating the enriched
portion of the simulation volume. The building
of the mask can be performed before the beginning of the simulation
and passed as additional IC or can be created in memory during the
simulation initialization. If the mask contains $m$-true values,
a shadow map of $m$ cell spectra, $S_{m}^{C}$, and luminosities, 
$L_m^C$, is allocated to store
the shapes and luminosities of the incoming packets: each time a packet enters a cell,
the mask is used to check whether the cell is an enriched one and
consequently $S_{m}^{C}$ and $L_m^C$ should be updated (see Step 2). 
The masking has been designed to optimise the performances of the code, but in principle
the SED and luminosity can be calculated in each cell of the computational
domain, if required.

\subsection{Step 2: radiative transfer simulation}

The next step (Step 2 in Fig.~\ref{fig:CRASH-PIPELINE})
consists in performing a RT simulation
which, in addition to the evaluation of 
($x_i,T$)$^C_c$ in all the cells of the domain, tracks also the SED
and luminosity of the ionising radiation in each of the $m$ 
metal enriched cells ($S,L$)$^C_m$. All the
above physical quantities are stored at times $t_k$, as already
mentioned above. 
The values of ($S,L$)$^C_m$ at $t_k$ are determined
adding up the contribution of all the incoming photon packets up to that time.
In practice, the code tracks all the multi-frequency packets entering the cell 
and calculates $L^C_m$ as the sum of the luminosities of all the packets,
while $S^C_m$ as the average SED in each frequency bin.

\subsection{Step 3: determination of the physical state of the metal enriched cells\label{sec:METALS-IN-DB}}

Finally, a search engine\footnote{We have implemented a Standard Query Language database using
one of the codes freely available to the scientific community. More details can be found in Appendix B}
looks for the \texttt{Cloudy} precomputed
configuration that best matches the values of H, He, metal number densities and 
$(x_{\rm HII},x_{{\rm HeII}},x_{{\rm HeIII}},S,L)_{m}^{C}(t_k)$.
The best fitting configuration then provides the ionisation fractions of the metal ions and the 
temperatures $T_{m}^{C3}$ of the gas in the metal enriched sub-domain
(Step 3 in Fig.~\ref{fig:CRASH-PIPELINE}).

If the matching criteria are not satisfied
(see below  and Appendix B for more details), the database is extended with additional
on-the-fly \texttt{Cloudy} runs.                                              
It is important to point out that Step 3 does not severely affect the basic
algorithm performances because it is confined only
to the $m$ enriched cells and a large number of {\tt Cloudy} calculations
are pre-computed and stored in the DB. The interested reader can find more details 
about the lookup strategy in Appendix B.

A correct computation of the temperature is not
trivial, because also in the absence of metals the temperatures predicted
by \texttt{CRASH} and \texttt{Cloudy} are not in
perfect agreement in the vicinity of the point sources (MCK09). More
generally, it has been shown that different approaches to the radiative
transfer do not always predict consistent temperatures in such regions
(\citealt{b68-Iliev2006b}). 
This means that every time a discrepancy between $T^C_m$ and $T^{C3}_m$
occurs it is important to understand
if this is due to the presence of metals or just to the differences in
the two codes. For these reasons we allow some flexibility in the matching criteria for the temperature and we define a temperature deviation $\Delta T_{m}^{i}$
as:

\begin{equation}
\Delta T_{m}^{i}=T_{m}^{C}-T_{m}^{C3},\label{eq:T_Correction}
\end{equation}
where $i=met$ refers to the deviation calculated for a gas contaminated
by metals, while $i=pris$ refers to a pristine gas. The difference
$\Delta T_{m}=\Delta T_{m}^{met}-\Delta T_{m}^{pris}$ is $\geq0$ by design
and it is due only to the effect of metals. In the enriched cells in
which $\Delta T_{m}$ is greater than some threshold value for the
maximum tolerated deviation, $T_{m}^{C}$ is replaced by $T_{m}^{C3}$.
We apply this correction (accounting for the metals feedback on the temperatures) 
when the $\Delta T_{m}$ is higher than $\sim$10\% $T_{m}^{C3}$ without metals. However the 
value adopted for the threshold depends on the physical problem being study and for 
this reason the criteria for the threshold can be defined at the beginning of each simulation.
Note that the temperature correction has some weak feedback on the
physical state of the gas also via its recombination coefficients, especially
for the helium component.

In realistic applications, the temperature provided by the hydrodynamics 
($T_{hydro}$) could be significantly higher than the temperature established by the RT. 
This generally happens in gas shocked regions where the metal ions are determined 
mainly by collisional ionisation rather than photo-ionisation.
Collisional ionisation is correctly handled by the pipeline at Step 2 for H and He, because \texttt{CRASH} 
selects $T=max(T^{C},T_{hydro})$ during its simulation initialisation process, maintaining the right temperature. 
Note that these regions are known as ICs, and a masking technique can be used to isolate them 
from the standard pipeline work-flow. 
The ionisation status of the metal component eventually present in these cells can be determined from 
\texttt{Cloudy} pre-computed tables, and by interpolating 
the gas number density, metallicity, temperature, as well as the ionisation fractions of hydrogen and helium.
For more details on this technique see \citet{b28-Oppenheimer2006} and references therein.

\section{Tests \label{sec:TESTS}}

In this Section we present three tests designed to establish the reliability
of the new code. The first test (Section~\ref{sub:StroemgrenTEST1})
concentrates on the standard Str\"omgren sphere case, albeit of a metal
enriched gas. The second test (Section$\,$\ref{sub:TEST-2}) investigates
the sensitivity of \texttt{CRASH3} to fluctuations of
the radiation field induced by different source types and tracked
by the many metal ionisation states. Finally, the third test (Section~\ref{sub:TEST3})
describes a more realistic physical configuration by using as ICs
those from a snapshot of a hydrodynamic simulation.

Hereafter the gas metallicity (or equivalently the metal mass fraction)
is defined as $Z_{g}=M_{Z}/M_{g}$, where $M_{Z}$ is the total mass
of the elements with atomic number higher than 2 and $M_{g}$ is the
total mass of the gas; the metal mass fraction in the Sun is set to 
$Z_{\odot}\approx0.0126$, according to the metal abundances relative
to hydrogen as reported in the \texttt{Cloudy} Hazy guide I and references
therein, and taking into account the 10 most abundant elements: H,
He, C, N, O, Ne, Si, Mg, S and Fe.

Unless otherwise stated, the metal feedback on the calculation of the gas
temperature is neglected and we use $k_f=100$.

\subsection{Test 1: Str\"omgren sphere with metals\label{sub:StroemgrenTEST1}}

We consider a configuration similar to the one in Test 2 of the Cosmological
Radiative Transfer Comparison Project \citep{b18-Iliev2006a}, but for the presence
of metals.

The simulation box has a co-moving side length of 6.6~kpc and it is
mapped onto a regular grid of $N_{c}^{3}=128^{3}$ cells. The gas is
assumed to be uniform and neutral at the initial temperature $T=100$~K,
with a number density $n_{gas}=0.1$ cm$^{-3}$ and a hydrogen (helium)
number fraction of 0.9 (0.1). Only one point source is considered
with coordinates (1,1,1), a black-body spectrum at temperature $T=10^{5}$~K
and  ionisation rate of $\dot{N}=10^{51}\textrm{phot s}^{-1}$ (i.e.
a luminosity $L\simeq5\cdot10^{40}\,\textrm{erg}\,\textrm{s}^{-1}$).
To ensure a good convergence of the MC scheme, the source radiation
field has been sampled by $2\cdot10^{8}$ photon packets. The redshift
of the simulation is set at $z=0$ and the simulation duration at
$t_{f}=5\cdot10^{8}\textrm{yrs}$, i.e. several times the recombination time characteristic of the simulated gas configuration. This choice assures that convergence is reached at the end of the simulation.
We uniformly contaminate the gas with carbon ($n_{{\rm C}}\simeq2.2\cdot10^{-7}$~cm$^{-3}$),
oxygen ($n_{{\rm O}}\simeq4.41\cdot10^{-7}$~cm$^{-3}$) and silicon ($n_{{\rm Si}}\simeq3.12\cdot10^{-8}$~cm$^{-3}$),
corresponding to $Z_{g}=6.45 \cdot10^{-3}Z_{\odot}$.
These numbers are obtained maintaining the metal abundances relative to hydrogen
mentioned above.
Finally, the source spectrum is sampled by 91 frequencies and it
extends to an energy $E_{max}=0.2$ keV, in order
to include the photo-ionisation edge of the ion O$\, \rm \scriptstyle VI$.

Although we contaminate the entire box with metals, it is possible
to significantly reduce the number of DB queries                   
taking advantage of the spherical geometry of the problem   
and assuming that each radial direction is equivalent. This is justified
as long as the number of photon packets used is large enough that
the fluctuations induced by the Monte Carlo sampling are negligible.
We then apply a spherical average of all the relevant physical quantities at each distance
$d$ (expressed in cell unit) from the ionising source, and we use these averaged values 
to search for the better-matching configuration in the DB. All the quantities 
discussed and displayed in this Test should be intended 
as explained above, i.e. as spherically averaged values.

\subsubsection{Pipeline convergence}
\label{sub:MetalCRASH-vs-CRASH2}

Before proceeding further with the analysis of the results, we discuss
the internal convergence of the pipeline. More specifically, we need to
assure that the ionisation fractions $x_{\rm HII}$, $x_{\rm HeII}$ and $x_{\rm HeIII}$
calculated in Step 2 and Step 3 of the pipeline are in agreement with some user-defined threshold. 
The same needs to be true for the temperature $T$ if the effect of metals is negligible. 
This would ensure that the physical configuration and energetic balance 
evaluated in Steps 2 and 3 are fully consistent. 
To this aim we run Test~1 in the absence of heavy elements.

\begin{figure}
\centering
\includegraphics[angle=-90,width=0.50\textwidth]{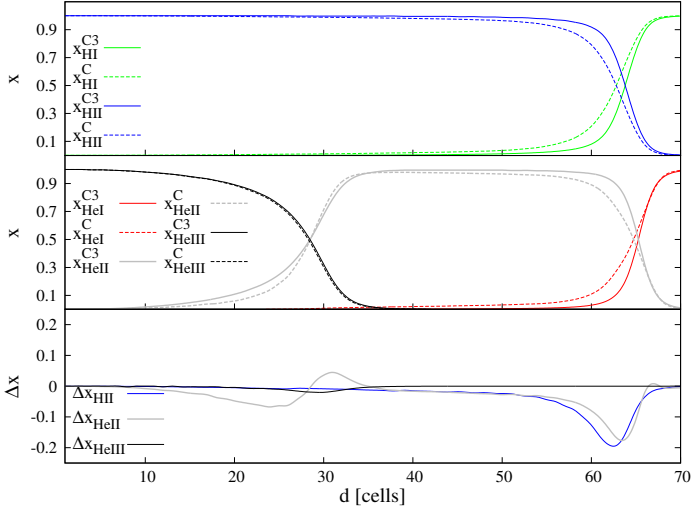}
\vspace{1truecm}
\caption{Convergence between
  Step 2 (dashed lines and variables with superscript
  $^{C}$) and Step 3 (solid lines and variables with
  superscript $^{C3}$) for Test 1 in the absence of metals (see text for details). The lines refer to the simulation time $t_{f}=5\cdot10^{8}$~yrs.
  At distances larger than 70 cells the gas is neutral and therefore
  it is not shown in the plots.
  \textbf{ Top panel:} profile of $x_{\textrm{HI}}$
  (green lines) and $x_{\textrm{HII}}$ (blue).
  \textbf{Middle panel:} same as above for $x_{\textrm{HeI}}$
  (red lines), $x_{\textrm{HeII}}$ (gray) and $x_{\textrm{HeIII}}$
  (black).
  \textbf{Bottom panel:} $\Delta x_i=x_i^C-x_i^{C3}$ for $i$=H$\, \rm \scriptstyle II$
  (blue line), He$\, \rm \scriptstyle II$ (gray) and He$\, \rm \scriptstyle III$ (black).}

\label{fig:Convergence-CRAS2-CRASH3}
\end{figure}

\begin{figure}
\centering
\includegraphics[angle=-90,width=0.50\textwidth]{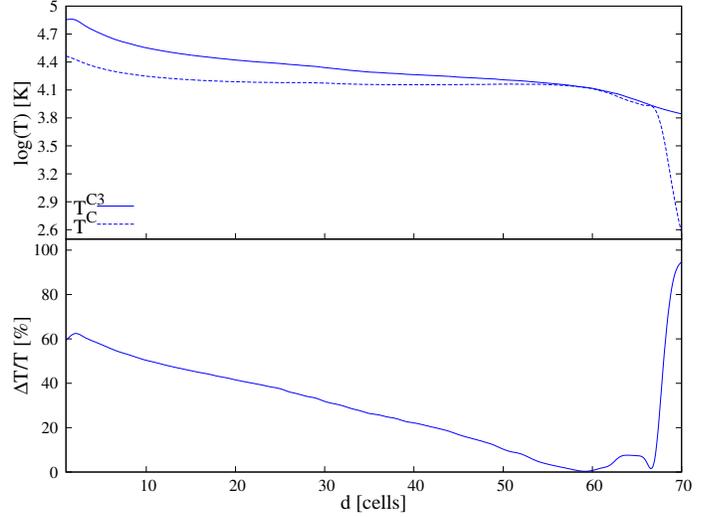}
\vspace{1truecm}
\caption{{\bf Top panel:} Temperature convergence between
  Step 2 ($T^C$, dashed line) and Step 3 ($T^{C3}$, solid line)                     
  for Test 1 in the absence of metals (see text for details). The lines
  refer to the simulation time $t_{f}=5\cdot10^{8}$~yrs.
  \textbf{Bottom panel:} relative difference $\Delta T/T=(T^{C3}-T^C)/T^{C3}$.}
\label{fig:Temperature-convergence-for}
\end{figure}

In Figure \ref{fig:Convergence-CRAS2-CRASH3} we show the profile
of $x_{{\rm HI}}$, $x_{{\rm HII}}$ (top panel), $x_{{\rm HeI}}$,
$x_{{\rm HeII}}$ and $x_{{\rm HeIII}}$ (middle panel) as evaluated
by Step 2 (dashed lines and variables with superscript $^{C}$) and
Step 3 (solid lines and variables with superscript $^{C3}$) 
at the time $t_{f}=5\cdot10^{8}\textrm{yrs}$.
The values of $x_{\textrm{HI}}$ and
$x_{\textrm{HII}}$ result in agreement
within $10^{-4}$ for $d<40$ cells ($\sim 2$ kpc) and within $10^{-2}$ up to the
ionisation front (I-front), identified as the location where the ionised
fraction drops below $\sim 0.8$. The agreement degrades
to $\sim18$ \% (with respect to the maximum absolute difference between ionisation fractions) in the two cells in which the
curves of \textbf{$x_{{\rm HI}}$} and \textbf{$x_{{\rm HII}}$} cross,
and then it restores up to $10^{-3}$.
Both codes predict the crossing in the same cell. 
A similar behaviour is shown in the middle panel for 
helium. A discrepancy of $\sim7$ \% in the $x_{\rm HeII}$ curves
is seen at a distance $d \sim 22$ cells, when the abundance of He$\,{\rm {\scriptstyle II}}$
starts to increase.
He$\,{\rm {\scriptstyle III}}$ results
in an even better agreement, up to $10^{-4}$. Reasonable accuracy
(less than $20$ \%) is also reached in the fronts of He$\,{\rm {\scriptstyle II}}$
and He$\,{\rm {\scriptstyle I}}$ where
both algorithms predict a similar shape. 
Because at the end of the HII region the intensity of the radiation has
dropped substantially (see following Section), sometimes {\tt Cloudy} does not
find a convergent solution in few iterations, and this results in a poorer agreement between Step 2
and 3.

The above differences are more evident in the bottom panel of the Figure, where
we show the absolute difference $\Delta x_i=x_i^C-x_i^{C3}$ for $i$=H$\, \rm \scriptstyle II$,
He$\, \rm \scriptstyle II$ and He$\, \rm \scriptstyle III$.

Despite the satisfying agreement in the \texttt{CRASH3} implementation,
some small discrepancies remain due to the differences between the
\texttt{CRASH} and \texttt{Cloudy} geometries and the numerical implementation
of the ionisation and energy equations (see Sec.~3). 
We found that a critical ingredient to reach an acceptable
convergence is to sample the source spectrum with a large number of
frequency bins (see details in the Test Section). This is necessary because the helium component is
very sensitive to this sampling, in particular in the vicinity of
the ionisation potential of  He$\,{\rm {\scriptstyle II}}$.

\begin{figure}
\centering
\includegraphics[angle=-90,width=0.50\textwidth]{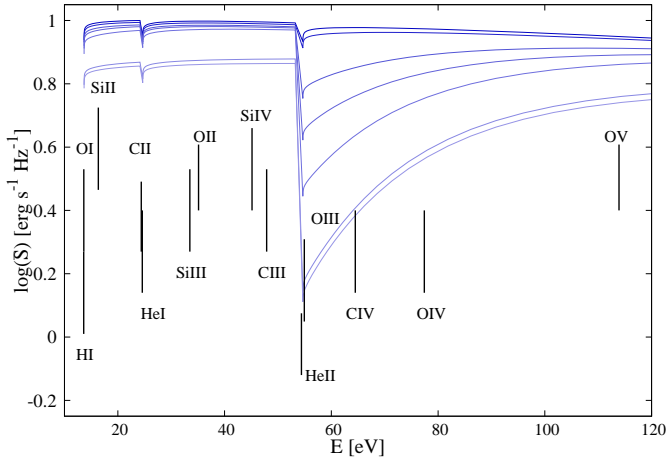}
\vspace{0.5truecm}
\caption{Normalized logarithm of
  the spherically averaged spectra. Lines
  refer to the spectra at a distance $d$ (from the top to the bottom) of 20,
  27, 38, 45, 55, 69 and 70 cells. The spectra correspond to
  a time $t_f=5 \cdot 10^{8}\textrm{yrs}$. As a reference, some ionisation potentials
  are also indicated as vertical lines.
  } 
\label{fig:SpectraVar1}
\end{figure}

The temperature radial profiles estimated in the two Steps are 
shown in Figure \ref{fig:Temperature-convergence-for} (top panel).
The curves present large discrepancies in the cells near the source, with
a relative variation $\Delta    T/T=(T^{C3}-T^C)/T^{C3}$ as high as 60 percent. 
The difference drops below 30 percent for $d>30$ cells. 
This is not reflected in the H and He profiles shown in  
Figure~\ref{fig:Convergence-CRAS2-CRASH3} because of the weak temperature
dependence in the gas recombination coefficients. Such a difference
has been already noticed and discussed in the \texttt{CRASH} vs \texttt{Cloudy}
comparison test in MCK09 and can be ascribed just to the different
implementation of the temperature estimate in the two codes.

Because \texttt{CRASH} updates the temperature (compared to its initial
value) only in those cells reached by ionising photons, outside the
HII region $T$ drops to the initial value of 100 K. On the other
hand, the temperature calculated by Step 3 is provided by \texttt{Cloudy}
and, as already mentioned above, all the regions where the intensity of
the radiation is too faint do not provide a convergent solution.
In the few cells in which \texttt{Cloudy} cannot provide a reliable 
estimate because of the too weak radiation field, the \texttt{CRASH} temperature $T^{C}$ 
is retained as better estimate of the physical temperature at the front.

\vspace{1truecm}

These convergence tests have been repeated using different ICs 
for the gas number density and the source luminosity.
 More specifically, we have run simulations on a grid of cases with
values $n_{gas}=1,\,0.1,\,0.01\,\textrm{cm}^{-3}$ and $\dot{N}=5\cdot10^{50},\,5\cdot10^{51}\textrm{phot s}^{-1}$.
It is found that, as the gas density decreases, the agreement improves
for the H species, while small discrepancies still remaining in the
He species. Such discrepancies are, on the other hand, always below
$20$ \% and remain limited to the small number of cells encompassing
the ionisation fronts. A similar improvement is observed also for the
gas temperature convergence.

Hereafter all the variables will refer to values calculated
at Step 3 and the superscript $^{C3}$ will be omitted to simplify
the notation.

\subsubsection{Ionisation field\label{sub:IonisationField}}

\begin{table}
\begin{centering}
\begin{tabular}{|c|c|c|c|c|c}
\hline
$E_{ion}\textrm{ [eV]}$ & H & He & C & O & Si\tabularnewline
\hline
\hline
$\textrm{E}_{\textrm{xI}}$ & \multicolumn{1}{c||}{13.598} & 24.587  & 11.260 & 13.618 & 8.152 \tabularnewline
\hline
$\textrm{E}_{\textrm{xII}}$ &  & 54.400  & 24.383  & 35.118  & 16.346 \tabularnewline
\hline
$\textrm{E}_{\textrm{xIII}}$ &  &  & 47.888  & 54.936  & 33.493 \tabularnewline
\hline
$\textrm{E}_{\textrm{xIV}}$ &  &  & 64.494  & 77.414  & 45.142 \tabularnewline
\hline
$\textrm{E}_{\textrm{xV}}$ &  &  & 392.090  & 113.900  & 166.770 \tabularnewline
\hline
$\textrm{E}_{\textrm{xVI}}$ &  &  & 489.997  & 138.121  & 205.060 \tabularnewline
\hline
\end{tabular}
\par\end{centering}
\caption{\label{tab:Ionization-Potentials}Ionisation potentials for H, He,
C, O and Si up to the ionisation level VI as used in {\tt Cloudy}.}
\end{table}

\begin{figure}
\centering
\includegraphics[angle=-90,width=0.50\textwidth]{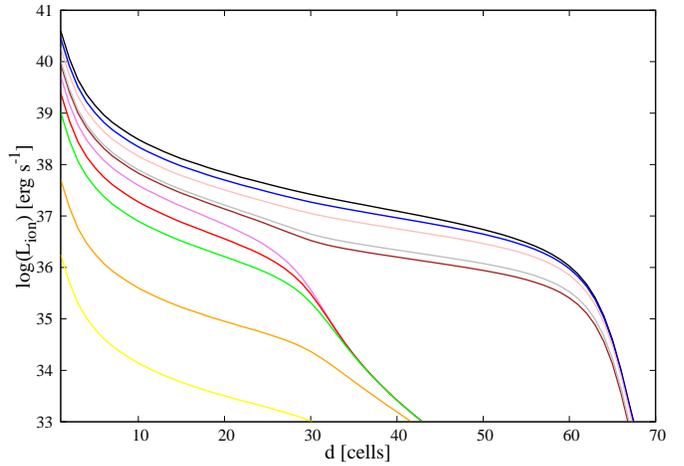}
\vspace{0.5truecm}
\caption{Logarithm of the photo-ionising luminosity, $L_{ion}$, 
  as function of $d$ at a time $t_f=5 \cdot 10^8$~yrs. 
  Different curves are calculated integrating all
  the photons above the ionisation threshold of (from the top to the bottom curve): 
  H$\, \rm \scriptstyle I$ (black line), 
  He$\, \rm \scriptstyle I$ (blue),
  Si$\, \rm \scriptstyle III$ (pink),
  Si$\, \rm \scriptstyle IV$ (gray),
  C$\, \rm \scriptstyle III$ (brown),
  He$\, \rm \scriptstyle II$ (violet),  
  C$\, \rm \scriptstyle IV$ (red), 
  O$\, \rm \scriptstyle IV$ (green), 
  O$\, \rm \scriptstyle V$ (orange), 
  O$\, \rm \scriptstyle VI$ (yellow).}
\label{fig:SpectraVar2}
\end{figure}

In Figure \ref{fig:SpectraVar1} we show how the spectral
shape of the ionising radiation field described in terms of its 
spherical average (obtained as described at the beginning of Sec.~5) 
changes with the distance $d$ as a result of geometrical dilution
and filtering. Each line refers to the simulation time $t_f=5 \cdot 10^{8}\textrm{yrs}$.

\begin{figure*}
\centering
\includegraphics[angle=-90,width=0.80\textwidth]{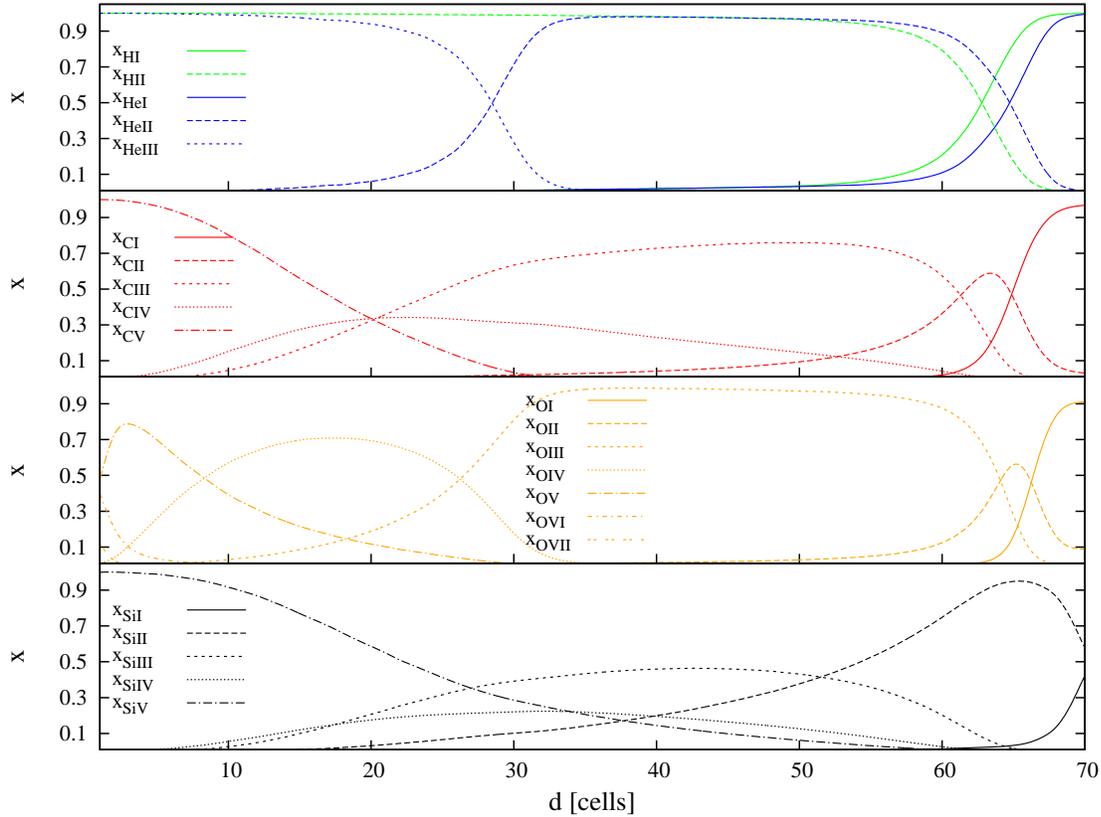}
\vspace{1truecm}
\caption{Fractions of the various components as a function
  of distance $d$ from the source in Test 1. The
  values are taken at the simulation time $t_f=5\cdot10^{8}\textrm{yrs}$.
  From the top panel to the bottom the species are: H (green lines)
  and He (blue); C (red); O (orange); and Si (black). In each panel the
  same ionisation states are represented by the same line-styles: solid
  lines refer to the neutral components (e.g. O$\, \rm \scriptstyle I$),
  long dashed to the first ionisation state (e.g. O$\, \rm \scriptstyle II$),
  short dashed to the second (e.g. O$\, \rm \scriptstyle III$),
  dotted to the third (e.g. O$\, \rm \scriptstyle IV$),
  long dashed-dotted to the fourth (e.g. O$\, \rm \scriptstyle V$),
  short dashed-dotted to O$\, \rm \scriptstyle VI$,
  and dashed-spaced to O$\, \rm \scriptstyle VII$.}
\label{fig:COSiIons}
\end{figure*}

The spectra shown are truncated at $E=120$~eV to have a better
visualisation of the most relevant line potentials. The upper curve
corresponds to a distance of $d=20$ cells, while the lower
to $d=70$ cells; at larger distances the intensity
of the radiation becomes too faint to solve the ionisation
equilibrium in every configuration at Step 3  of the \texttt{CRASH3} pipeline.  
Because the medium is fully transparent in H and He up to a distance $d=20$ cells, we
have normalized the spectra in the Figure to the maximum value of the spectrum at $d=20$ cells.
This choice allows a better visualisation of the shaping effects by the partially ionised 
regions. 
The ionisation potentials of the metals
enriching the box (see Table~\ref{tab:Ionization-Potentials})
are also shown as reference even if, by design,
metals do not contribute to the filtering of the ionising radiation.
The absorption due to H$\,{\rm {\scriptstyle I}}$, He$\,{\rm {\scriptstyle I}}$
and He$\,{\rm {\scriptstyle II}}$ is clearly visible in correspondence
of the respective ionisation potential, i.e. 13.6~eV, 24.6~eV and
54.4~eV. 

Figure \ref{fig:SpectraVar2} shows the total photo-ionising
luminosity, $L_{ion}$, available for ionisation of some reference species
as a function of $d$ at a time $t_f=5 \cdot 10^{8}$~yrs.
$L_{ion}$ is defined as:
\begin{equation}
L_{ion}\left(d\right)={h_{p}}^{-1}\intop_{E_{ion}}^{E_{max}} S(E,d) \, dE,\label{eq:L-per-Ion}
\end{equation}
 where $E_{ion}$ is the ionisation potential of the
species considered (see Table \ref{tab:Ionization-Potentials} for
reference) and $h_{p}$ is the Planck constant.

Because the spectrum includes only photons with $E>13.6\,\textrm{eV}$,
$L_{ion}$ is an underestimate for those elements with an ionisation
potential below $13.6\,\textrm{eV}$, i.e. carbon and silicon.
Note that $L_{\rm OI}$, $L_{\textrm{CII}}$ and $L_{\rm OIII}$           
overlap respectively with $L_{\rm HI}$ (black line), $L_{\textrm{HeI}}$ (blue) and $L_{\rm HeII}$ (violet) because of
the very similar ionisation potential. If we increased the frequency
resolution of the spectrum the curves would show some small difference.
This would be at the expense of the computational time without significant
advantages in the accuracy of the metal ionisation state. For this
reason we do not further increase the frequency resolution.
Notice that all the luminosities decrease smoothly with $d$; this
is consistent with the behaviour expected in an HII region, as already
reported and extensively commented in \citet{b24-Maselli2003} and MCK09 for the
hydrogen and helium components.

\begin{figure*}
\centering
\includegraphics[angle=-90,width=0.80\textwidth]{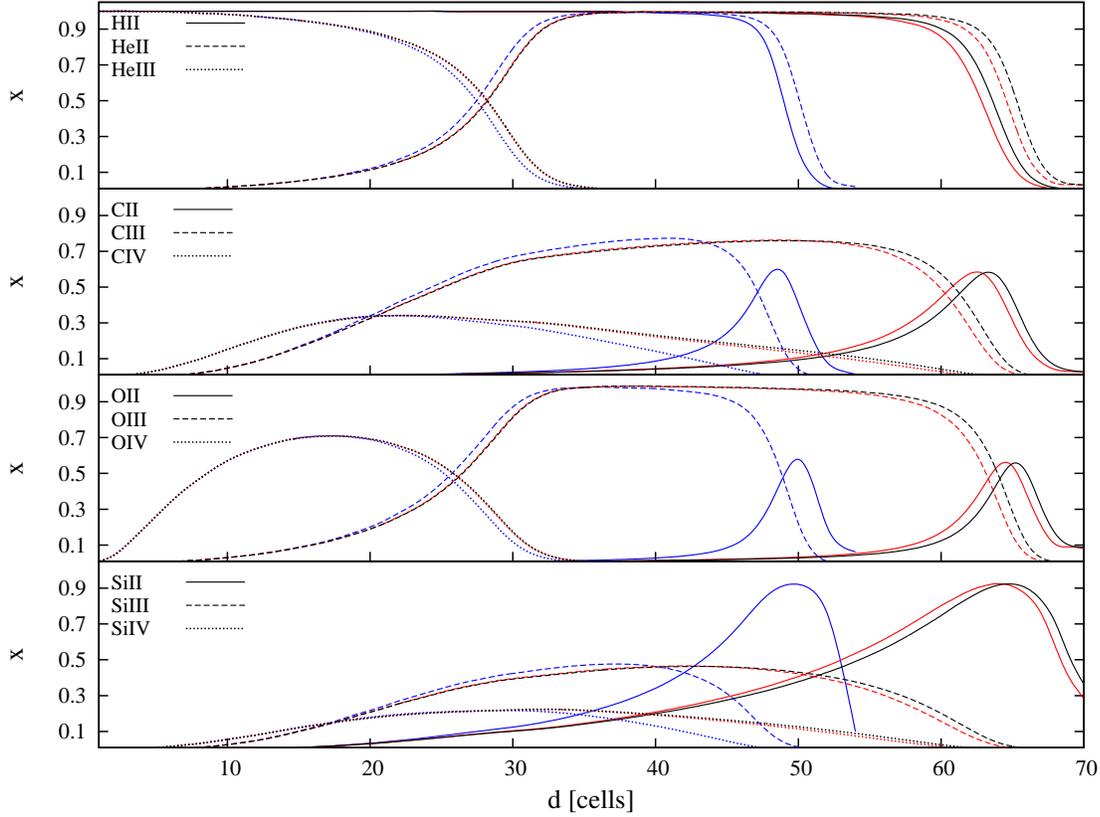}                
\vspace{1.0truecm}
\caption{Fractions of the various components as a function
  of the distance $d$ from the source in Test~1. From the top panel to the bottom the species are:
  H and He; C; O; and Si. In each panel colors refer to
  different simulation time: $t=10^6$ (blue), $10^7$ (red) and $5 \cdot 10^8$~yrs (black).}
\label{fig:TEvolution}
\end{figure*}

\subsubsection{Metal ionisation states\label{sub:MetalIonizedStates}}

We now analyse the behaviour of the metal ionisation states, plotted
in Figure \ref{fig:COSiIons}. Before discussing the details
of the Figure, it is necessary to point out that the balance among
the different ions is primarily established by the relative values
of their ionisation potentials and of their recombination
coefficients (see \citealt{b11-Ferland2003}). It also depends on the spectral distribution
of the radiation field and its variations with the distance from the
source, as induced by the radiative transfer effects. The complex
interplay between these numerous processes makes the interpretation
of the results non trivial; despite this, some trends have a straightforward
explanation.

Because of the small amount of metals included in this test, we expect
their impact on the evolution of H and He to be negligible. This is confirmed 
by a comparison of the curves in the top panel of the
Figure to the corresponding curves in Figure \ref{fig:Convergence-CRAS2-CRASH3},
which are obtained without metals. The maximum difference is $\sim7$ \% across the I-front
of H$\, \rm \scriptstyle II$.
Additional effects on the ionisation fractions induced
by an increase in the gas metallicity will be investigated in Section
\ref{sub:Test1MetalFeedback}. 

We now turn to analyse the behaviour of carbon (second panel from the top).
For $d<30$ cells, C is in the form of C$\, \rm \scriptstyle III$,
C$\, \rm \scriptstyle IV$ and C$\, \rm \scriptstyle V$, with a predominance of
the latter close to the source. This high
ionisation level is obtained from a combination of collisional ionisation
and photo-ionisation. The evolution of C$\, \rm \scriptstyle III$
is very similar to that of He$\, \rm \scriptstyle II$ because of
the similar ionisation potentials (see Fig.~\ref{fig:SpectraVar1}).
The abundance of C$\, \rm \scriptstyle IV$
is dictated by the evolution of C$\, \rm \scriptstyle III$
and C$\, \rm \scriptstyle V$, and their relative recombination
coefficients.
C$\, \rm \scriptstyle IV$ is present throughout the entire HII region,
with $x_{\textrm{CIV}}$ being always below 30 percent. For
$d\gtrsim30$ cells $x_{\textrm{CV}}$ goes to zero because only few
C$\, \rm \scriptstyle IV$ ionising photons are available (see Fig.~\ref{fig:SpectraVar2}
as a reference). The ionisation
potential of C$\, \rm \scriptstyle V$ is outside our frequency
range (see Table \ref{tab:Ionization-Potentials}) and thus higher
ionisation states are not present. 
Because of the paucity of photons with $E>E_{\textrm{CIII}}$, at $d>30$ cells only C$\, \rm \scriptstyle III$
is present in large quantities with $x_{\textrm{CIII}}\sim70\%$.
At $d\sim60$ cells, similarly to what happens to H$\, \rm \scriptstyle II$
and He$\, \rm \scriptstyle II$, also C$\, \rm \scriptstyle III$ declines
and C$\, \rm \scriptstyle II$ dominates. Finally, outside 
the HII region, only C$\, \rm \scriptstyle I$ is present.

In the third panel from the top the ions of the oxygen are shown. The
ionisation potential for O$\, \rm \scriptstyle VI$ is the highest
photo-ionising energy available in the adopted spectrum. A very small
fraction of O$\, \rm \scriptstyle VII$ is in fact present in few
cells around the source. It should be noticed that collisional ionisation contributes 
to this fraction for $\sim$ 10 \% at $T\geq7\cdot10^{4}$~K, which is
present when $d<3$ cells.
The presence of O$\, \rm \scriptstyle VI$, is more evident but
it is consistently limited to the inner region of the
ionised sphere and decreases rapidly with the distance from the source
in favour of lower ionisation levels.                                     
For $d\gtrsim30$ cells O$\, \rm \scriptstyle III$ dominates the
ionisation balance until it drops in favour of O$\, \rm \scriptstyle II$, roughly at the end of the HII region.
As for the other species, outside the HII region only O$\, \rm \scriptstyle I$
is present.

\begin{figure*}
\includegraphics[angle=-90,width=0.80\textwidth]{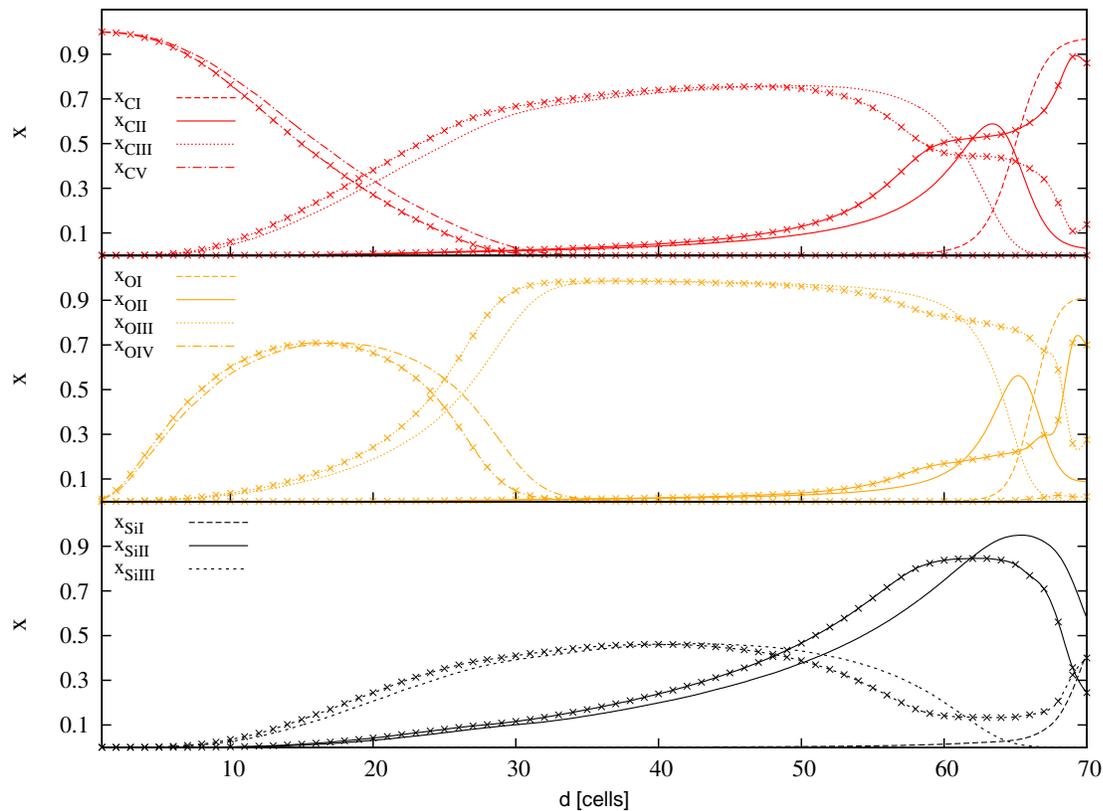}              
\vspace{1.0truecm}
\caption{Fractions of C ({\bf upper panel}), O ({\bf middle}) and Si ({\bf lower}) ions
  as a function of the distance $d$ from the source in Test 1.
  In each panel, lines with crosses refer to Test~1 with the source
  spectrum extending to a lower ionising energy of $E_{\textrm{SiI}}$,
  rather than $E_{\rm HI}$. Note that the neutral fractions in the former case
  are equal to zero in the range of cells shown here.
  }
\label{fig:IRExt}
\end{figure*}

In the bottom panel the behaviour of silicon is reported.
Si$\, \rm \scriptstyle V$ completely dominates the inner region
of the Str\"omgren sphere, with a long tail extending
to $d\sim55$ cells, where it is in equilibrium
with lower ionisation states. In the central region many ions are in
equilibrium with a low ionisation fraction. Si$\, \rm \scriptstyle III$
reaches a maximum fraction of $\sim 0.4$ at the center of the HII region.
Si$\, \rm \scriptstyle II$ dominates at $d>52$ cells.             
The abundance of Si$\, \rm \scriptstyle I$ in the outskirts of the HII
region is much lower than the one of e.g. C$\, \rm \scriptstyle I$
because of the lower number density of Si and the much larger
cross section of Si$\, \rm \scriptstyle I$ at the resonance 
(see Fig. 13.2 of \citealt{b88-Draine2011}).

In general, it is possible to say that in the vicinity of the source
the most abundant species are those with the higher ionisation state
compatible with the maximum potential in the spectrum, i.e. H$\,{\rm {\scriptstyle II}}$,
He$\,{\rm {\scriptstyle III}}$, C$\,{\rm {\scriptstyle V}}$,
O$\,{\rm {\scriptstyle V}}$ and Si$\,{\rm {\scriptstyle V}}$.
Despite $E_{{\rm OV}}$, $E_{{\rm OVI}}$ and $E_{{\rm SiV}}$ being
covered by the spectrum, the abundance of photons at these energies
is so low that $x_{{\rm OVI}}$, $x_{{\rm OVII}}$ and $x_{\textrm{SiVI}}$
are negligible. As the distance increases, the luminosity available
for ionisation decreases, in particular for ions with high ionisation
potentials (see Fig.~\ref{fig:SpectraVar2}). This is reflected by
the decrease of the abundance of these highly ionisation states and the
predominance of lower ionisation states (e.g. He$\,{\rm {\scriptstyle II}}$,
C$\,{\rm {\scriptstyle III}}$, O$\,{\rm {\scriptstyle III}}$
and Si$\,{\rm {\scriptstyle III}}$). Species like Si$\,{\rm {\scriptstyle IV}}$
and C$\,{\rm {\scriptstyle IV}}$ are always present, although they
are not dominant, because the spectrum of the ionising radiation maintains
energies higher than $E_{{\rm SiIII}}$ and $E_{{\rm CIII}}$ throughout
the HII region (see Fig.~\ref{fig:SpectraVar2}). At even larger
distances ($d>60$ cells) the dominant species are typically singly
ionised metals and the neutral components, due to the absence of ionizing radiation. 

While the discussion above refers to the final gas configuration,
in Figure \ref{fig:TEvolution} we show some reference species at different
simulation times, i.e. $t=10^6$, $10^7$ and $5 \cdot 10^8$~yrs.
It should be kept in mind that in this case no metal feedback is
included in the calculation (see Sec.~\ref{sub:Test1MetalFeedback}) and thus the profiles
of the metal species at each time are independent from each other.
While the shape of each species changes substantially between  
$10^6$ and $10^7$~yrs, for $t \simgt 10^7$~yrs they have almost
converged. This is especially true for the metal ionisation potentials
larger than the third and for He$\,{\rm {\scriptstyle III}}$.
Overall, the comments made for Figure~\ref{fig:COSiIons} at $t_f=5 \cdot 10^8$~yrs
apply also at earlier times.

\subsubsection{Effect of photons with $E<13.6$~eV}
\label{sec:lowerenergy}

In this Section we discuss the impact of photons with $E<13.6$~eV in the evaluation 
of metal ionisation states. Because the ionisation potential of
C$\, \rm \scriptstyle I$ and Si$\, \rm \scriptstyle I$ is below $E_{\textrm{HI}}$, 
i.e. in a range which is not covered by \texttt{CRASH} and where the spectrum 
is set by default to zero, this means that photo-ionisation by photons with $E<13.6$~eV is
neglected, resulting in a systematic underestimate
of $x_{\textrm{CII}}$ and $x_{\textrm{SiII}}$ as visible in the outer region
of the Str\"omgren sphere.

To investigate this limitation of our pipeline, we have extended the spectrum
to a lower energy of $E_{\textrm{SiI}}=8.152$~eV. It should
be kept in mind, though, that in the range $E_{{\rm SiI}}$-$E_{{\rm HI}}$
the luminosity of the spectrum is overestimated because the absorption
by metals (the only species contributing to the optical depth in this
energy range) is not accounted for. 
This means that the change in the
spectrum at these frequencies is due only to geometrical dilution. 
The approximation is more severe
beyond the He$\, \rm \scriptstyle II$ front (see Figure
\ref{fig:COSiIons}), where the only surviving radiation has energy below
$E_{\textrm{HI}}$ and the optical depth of the medium is dominated
by the metals\footnote{In realistic configurations other absorbers like dust and molecules 
could play a dominant role in establishing the optical depth of the medium at these frequencies, but
these are not accounted for in \texttt{CRASH3}.}.

In Figure \ref{fig:IRExt} we summarise the results of this run by showing 
the evolution of different C, O and Si ions. Lines with crosses refer to the case
in which the source spectrum extends below 13.6~eV.
In the inner part of the HII region the agreement between the curves with and
without the inclusion of the low energy tail is excellent for the neutral
fractions and the first ionisation state, while it gets worse for higher
ionisation states. The reason is that, when the spectrum is extended
below 13.6~eV additional physical processes become relevant, in particular line
emission from collisional excitation. The most prominent case is the collisional
excitation of the O$\, \rm \scriptstyle IV$ line. While the behaviour of
$x_{\rm OV}$ is the same in both cases (and for this reason it is not reported
here), in the presence of non H$\, \rm \scriptstyle I$-ionising radiation O$\, \rm \scriptstyle IV$
recombines more quickly to O$\, \rm \scriptstyle III$, inducing the discrepancies
observed between $\sim$ 20-35 cells. Something similar happens to C$\, \rm \scriptstyle V$
and C$\, \rm \scriptstyle III$, while C$\, \rm \scriptstyle IV$ is not affected 
because it is not a dominant species at these distances, as shown in Figure~\ref{fig:COSiIons}.
For the Si the differences are smaller.
The picture in the outskirts of the HII region instead changes completely:
not only do $x_{\textrm{CII}}$,  $x_{\rm OII}$ and $x_{\textrm{SiII}}$ extend
outside the HII region, while the corresponding neutral components
disappear, but also the higher ionisation states are affected. This results
from a combination of the photo-ionisation due to the photons with energies below
13.6~eV and the additional physical processes mentioned above. 
At these distances the former effect has a larger impact than in
the vicinity of the source, where ionisation is dominated by photons with
energies above 13.6~eV.
It should be noted that outside the HII regions, or, more correctly, when the
absorption of photons is not dominated any more by H and He, our assumption of
neglecting the metal contribution to the optical depth fails and the metal
ionisation states are not calculated correctly any more\footnote{Other important processes playing a relevant role in the outskirts of the HII region, as e.g. $O^{+} + H \longleftrightarrow O + H^{+}$, have been deactivated in {\tt Cloudy} by disabling the charge transfer. More references on the subject can be found in \cite{b88-Draine2011}.}.

\subsubsection{Feedback by metals\label{sub:Test1MetalFeedback}}

In this Section we investigate the effect of metals on the evaluation of
the gas temperature. In the Appendix A the convergence of this feedback with
respect to the choice of $k_f$ is investigated. This is done by calculating $T$ at the time
$t_f=5 \cdot 10^8$~yrs in the standard configuration of Test~1 and
then changing the metallicity of the gas.                                   

\begin{figure}
\includegraphics[angle=-90,width=0.50\textwidth]{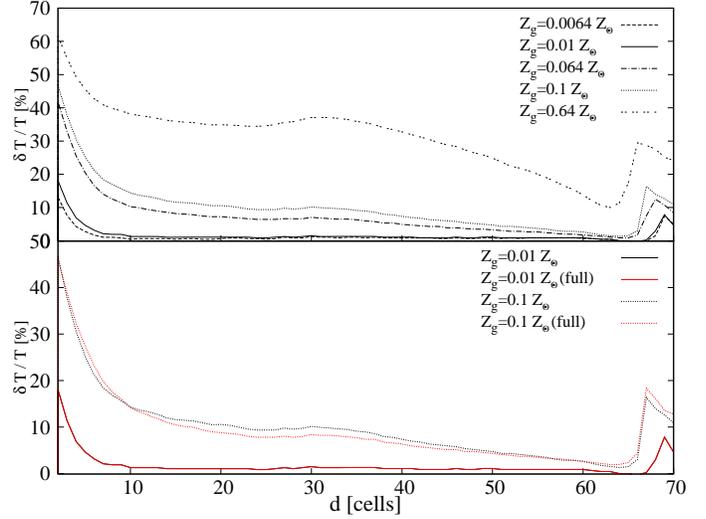}              
\vspace{0.5truecm}
\caption{$\Delta T/T=(T(0)-T(Z_g))/T(0)$ as function of the distance $d$ from the 
  source in Test~1 at the  
  simulation time $t_{f}=5\cdot10^{8}$~yrs (see text for more details).
  {\bf Upper panel:} the curves refer to a gas enriched by C, O and Si, and a metallicity $Z_{g}$ of: 
  $0.0064Z_{\odot}$
  (dashed line, reference value), $0.01Z_{\odot}$ (solid), $0.064Z_{\odot}$ (dashed-dotted), 
  $0.1Z_{\odot}$ (dotted) and $0.64Z_{\odot}$ (dashed-spaced).
  {\bf Lower panel:} the curves refer to a gas enriched only with C, O and Si (black lines)
  or with the ten most abundant elements in the solar composition (red). For both cases the gas
  metallicity $Z_g$ is $0.01Z_{\odot}$ (solid lines) and $0.1Z_{\odot}$ (dotted).}

\label{fig:TCorrectionMetallicity}
\end{figure}

\begin{figure}
\includegraphics[angle=-90,width=0.50\textwidth]{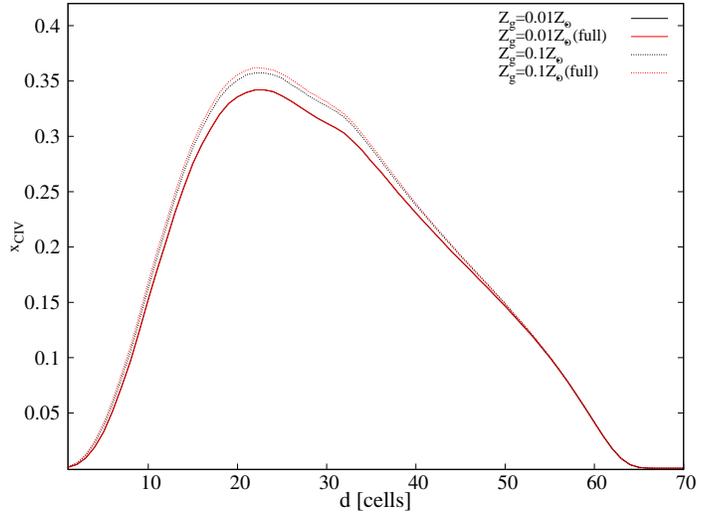}       
\vspace{0.5truecm}
\caption{Fraction of C$\, \rm \scriptstyle IV$
  as function of the distance $d$ from the source in Test 1 at the simulation
  time $t_f=5 \cdot 10^8$~yrs. The curves refer to a gas enriched only with
  C, O and Si (black lines) or with the ten most abundant elements in the 
  solar composition (red). For both cases the gas metallicity $Z_g$ is
  $0.01 Z_{\odot}$ (solid lines) and $0.1 Z_{\odot}$ (dotted).
}
\label{fig:MetalCompositionFeedbackOnIons}
\end{figure}

Test 1 has been repeated changing the gas metallicity $Z_{g}$, while maintaining the relative
abundance of C, O and Si. 
In the upper panel of Figure~\ref{fig:TCorrectionMetallicity} we show the results in terms of:
\begin{equation}
\Delta T/T\equiv\frac{T\left(0\right)-T\left(Z_{g}\right)}{T\left(0\right)},\label{eq:T_Metallicity_ratio}
\end{equation}
 where $T\left(0\right)$ is the value of the temperature
relative to a configuration with $Z_{g}=0$.
The reference case ($Z_g=0.0064Z_{\odot}$, dashed line) does not show any significant
metal cooling, with the exception of the region near to the source ($d<5$
cells) where recombination and re-emission of high ionisation states of C, O and Si
is more significant, inducing an average  $\Delta T/T\sim10\%$. Temperature
deviations at the He$\, \rm \scriptstyle II$ I-front ($d>65$ cells),
where the ionising radiation is very faint, are also present, with
$\Delta T/T<10\%$. 
Increasing the metallicity to one percent solar (solid line)        
does not change the results. Only at $Z_{g}=0.064Z_{\odot}$
(dashed-dotted line) some cooling is visible at each distance. In few
cells near the source $\Delta T/T$ is as high as 40$\%$, while it              
remains below 10 percent at $d>10$ cells. 
The effect of metal cooling starts to be very significant for 
$Z_{g}\geqslant0.64Z_{\odot}$ (dashed-spaced line), with  a
$\Delta T/T\sim60\%$ in the inner part of the HII region and $\sim 15-20\%$ also
in its outer part. 

Even if it is not shown in Figure~\ref{fig:TCorrectionMetallicity}, we have
computed the solar and super-solar cases ($Z_{g}\geq1Z_{\odot}$), finding a deviation
of $\Delta T/T>50\%$ in the inner region, and values exceeding 40 percent
up to $d\sim45$ cells. These results confirm the dominant role played by the
metal cooling in this metallicity range. These cases though should be considered only as indicative because in this metallicity regime the metal contribution
to the absorption cannot be neglected and the assumptions of our method
fail.

As a second step we have investigated the dependence of our results
on the gas composition; this is shown in the bottom panel of Figure
\ref{fig:TCorrectionMetallicity}. More specifically, while the black
lines are the equivalent of the ones in the upper panel, the red lines
have been obtained including the 10 most abundant elements in the
solar composition. Note that the red and black lines are not distinguishable 
in the reference metallicity case. In both cases the number densities of metals are such
that their abundance relative to H is the same as those in the solar
composition. It is clear that, at a fixed metallicity, 
the contribution of C, O and Si to the cooling
is dominant compared to other elements, such
as N, Ne, Mg, S and Fe. We have
thus neglected such elements in all further tests.
It should be noted that just adding the contribution of N, Ne, Mg, S and Fe
to the gas metallicity (without keeping it constant) would bring $Z_g$ from
e.g. $0.0064Z_{\odot}$ to $0.01Z_{\odot}$, and $0.064Z_{\odot}$ to $0.1Z_{\odot}$.

In Figure \ref{fig:MetalCompositionFeedbackOnIons} we finally report
the effect of a varying metallicity and chemical composition on
the ionisation fraction of C$\, \rm \scriptstyle IV$. 
It is clear that it is safe to neglect metals other than C, O and Si
as their impact on the global ionisation equilibrium is limited at a maximum of few percent for $Z_g \gsim 0.1 Z_{\odot}$,
while it is close to zero for lower metallicity. A similar effect is observed
for all the other relevant ions.
On the other hand, the dependence on the gas metallicity $Z_g$ is higher and it changes for the various species. In some cases, like the one shown in the Figure
(as well as e.g. O$\, \rm \scriptstyle III$ and Si$\, \rm \scriptstyle IV$), the ionisation fraction of the species increases with increasing metallicity, while in others (as e.g. 
C$\, \rm \scriptstyle III$, O$\, \rm \scriptstyle V$ and Si$\, \rm \scriptstyle IV$) the trend
is reversed. It is beyond the scope of this paper though to discuss this issue in more
details.

\subsection{Test 2: metal fluctuations in the overlap of two HII regions}
\label{sub:TEST-2}

\begin{figure}
\centering
\includegraphics[scale=0.67]{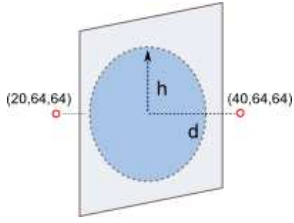}
\caption{Sketch of the geometrical set-up used for Test 2.}
\label{fig:Test2Sketch}
\end{figure}

Here we study the behaviour of metals in the HII region overlap produced by 
two point sources located
in cells $(20,64,64)$ and $(40,64,64)$. This geometrical set-up is
sketched in Figure \ref{fig:Test2Sketch} and it is designed to investigate
the sensitivity of our method to small changes in the source characteristics.
This is done by varying the source ionising rates, $\dot{N}_{1}$ and $\dot{N}_{2}$, and
their associated black-body spectral temperatures, $T_{1}$ and
$T_{2}$.
All the other numbers are the same as those in Test~1, with
the exception of the gas number density which is $n_{gas}=1$~cm$^{-3}$, 
to obtain a sharper overlap profile.
Because in this test we are interested in studying the behaviour of
metals only in the overlap region, we concentrate on the plane corresponding
to $x=30$ (blue circle in Fig.~\ref{fig:Test2Sketch}).

\subsubsection{Reference case\label{sub:Test2Reference}}

\begin{figure}
\centering
\includegraphics[angle=-90,width=0.50\textwidth]{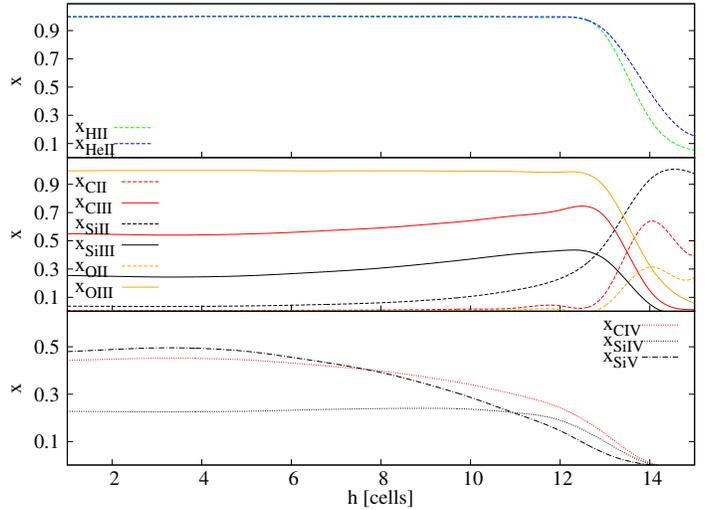}
\vspace{1.0truecm}
\caption{Ionisation fractions as function of $h$, evaluated in the plane equidistant
  from the two sources ($x=30$) of Test~2 in the reference case.
  From top to bottom the panels refer to:
  $x_{\textrm{HII}}$ (green dashed) and $x_{\textrm{HeII}}$ (blue
  dashed); $x_{\textrm{CII}}$ (red dashed), $x_{\textrm{CIII}}$ (red solid), 
  $x_{\textrm{SiII}}$ (black dashed), $x_{\textrm{SiIII}}$ (black solid),
  $x_{\textrm{OII}}$ (orange dashed) and x$_{\textrm{OIII}}$ (orange solid);
  $x_{\textrm{CIV}}$ (red dotted), $x_{\textrm{SiIV}}$ (black dotted) and $x_{\textrm{SiV}}$
  (black dashed-dotted). 
  }
\label{fig:Behaviour-LvarTfix}
\end{figure}

We first describe the set-up used as reference case, with
two identical sources with $\dot{N}_{1}=\dot{N}_{2}=9 \cdot 10^{51}$~phot~s$^{-1}$
and $T_{1}=T_{2}=10^{5}$~K.
Similarly to Test~1, here we show the results by averaging the
input physical quantities on all the cells of the plane at the same distance
$h$ from cell (30,64,64) (see Fig.~\ref{fig:Test2Sketch}).
The results refer to the final time $t_f=5 \cdot 10^8$~yrs.
Figure  \ref{fig:Behaviour-LvarTfix} shows the profiles of the ionization fraction of some selected ions.
 
The profile of the H and He species is very similar to that 
of a single Str\"omgren sphere (see Test 1), with the exception of 
$x_{\textrm{HeIII}}$, which is always confined in regions very close
to the sources and thus is not shown. 

In the middle panel C$\, \rm \scriptstyle II$,
C$\, \rm \scriptstyle III$, Si$\, \rm \scriptstyle II$, Si$\, \rm \scriptstyle III$,
O$\, \rm \scriptstyle II$ and O$\, \rm \scriptstyle III$ are
shown together because they trace the external regions of the two
overlapping Str\"omgren spheres (see Fig.~\ref{fig:COSiIons}). 
C$\, \rm \scriptstyle III$, O$\, \rm \scriptstyle III$ and Si$\, \rm \scriptstyle III$
are present for $h<12$ cells with different ionisation fractions
($x_{\textrm{OIII}}\sim1$, $x_{\textrm{CIII}}\sim0.5$ and $x_{\textrm{SiIII}}\sim0.3$),
indicating that these ions have a different sensitivity to the ionising
field; this is also in qualitative agreement with the relative trends
noticed in Test 1 (see Fig.~\ref{fig:COSiIons}).

In the lower panel $x_{\rm CIV}$, $x_{\rm SiIV}$ and 
$x_{\rm SiV}$ are shown. Similarly to the species analysed in the
previous panel, they present an almost constant value throughout
the fully ionised region, with the exception of $x_{\rm SiV}$, which
starts declining earlier, in favour of lower ionisation states.
The absence of O$\, \rm \scriptstyle IV$,
O$\, \rm \scriptstyle V$ and C$\, \rm \scriptstyle V$ (which reach
an ionisation fraction of only a few percent) is consistent with
the absence of He$\, \rm \scriptstyle III$.
Compared to the corresponding species in Test~1, here the ionisation
fractions are higher due to the larger ionisation rate.

\subsubsection{Variations in the source ionisation rates}

\begin{figure}
\centering
\includegraphics[angle=-90,width=0.50\textwidth]{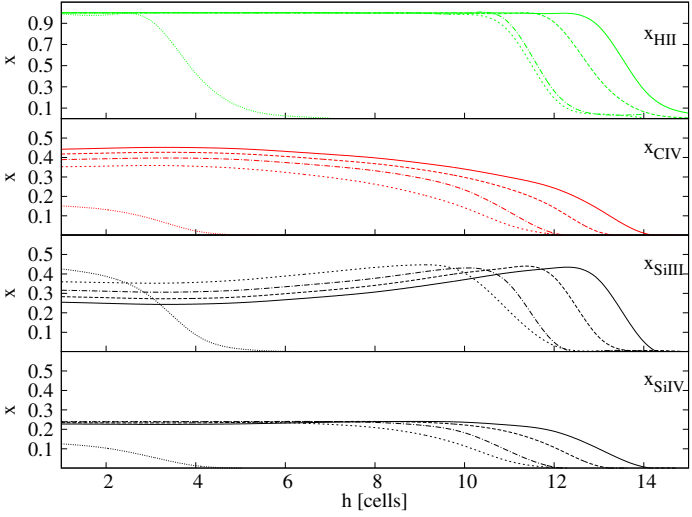}
\vspace{1.0truecm}
\caption{Ionisation fractions as function of $h$, evaluated in the plane equidistant
  from the two sources ($x=30$) of Test~2 when the source ionisation rate is
  changed compared to the reference case (see text for details). 
  From the top to the bottom the panels refer to $x_{\textrm{HII}}$,
  $x_{\textrm{CIV}}$, $x_{\textrm{SiIII}}$ and $x_{\textrm{SiIV}}$.
  In all panels solid lines refer to the reference case with 
  $\dot{N}_{1}=\dot{N}_{2}=9 \cdot 10^{51}$~phot~s$^{-1}$; dashed to 
  $\dot{N}_{1}=9 \cdot 10^{51}$~phot~s$^{-1}$ and 
  $\dot{N}_{2}=7 \cdot 10^{51}$~phot~s$^{-1}$; short-dashed to
  $\dot{N}_{1}=9 \cdot 10^{51}$~phot~s$^{-1}$ and 
  $\dot{N}_{2}=3 \cdot 10^{51}$~phot~s$^{-1}$; dashed-dotted to
  $\dot{N}_{1}=\dot{N}_{2}=7 \cdot 10^{51}$~phot~s$^{-1}$; dotted to
  $\dot{N}_{1}=\dot{N}_{2}=3 \cdot 10^{51}$~phot~s$^{-1}$.
}
\label{fig:Behaviour-L1varL2fixTfix}
\end{figure}

In this Section we discuss the variations in the results induced by changes
in the ionisation rates. These are shown in Figure \ref{fig:Behaviour-L1varL2fixTfix}
for some reference species.

First we have decreased the ionisation rates of the two sources simultaneously, 
maintaining the symmetry of the problem, i.e. using $\dot{N}_1=\dot{N}_2=7\cdot10^{51}$~phot~s$^{-1}$
and $3\cdot10^{51}$~phot~s$^{-1}$, the latter being the minimum ionising rate
allowing for an overlap of the two HII regions. 
We have then decreased the ionisation rate of only one source, while maintaining
$\dot{N}_{1}=9 \cdot 10^{51}$~phot~s$^{-1}$.
As expected, the extent of the fully ionised region (upper panel) decreases
with decreasing total ionisation rate.
A similar trend is observed also in the profile of the other species, which
changes smoothly with decreasing ionisation rate. Compared to the behaviour
of $x_{\rm HII}$ though, these differ in that also the amount of ionisation
decreases. Remarkably, $x_{\textrm{SiIV}}$
seems to be insensitive to small variations of the ionisation rates, at least
in the inner parts of the HII region. 
This suggests that other species would
be more suitable to give indications on the characteristics of the sources.

By comparing qualitatively the results with those in Figure~\ref{fig:COSiIons}
we can conclude that \texttt{CRASH3}
produces the predictable behaviour for the ions also when more than one source
is present and that it is
sensitive to small changes in the source ionisation rates.
This is very important for realistic applications of the code.

\begin{figure}
\centering
\includegraphics[angle=-90,width=0.50\textwidth]{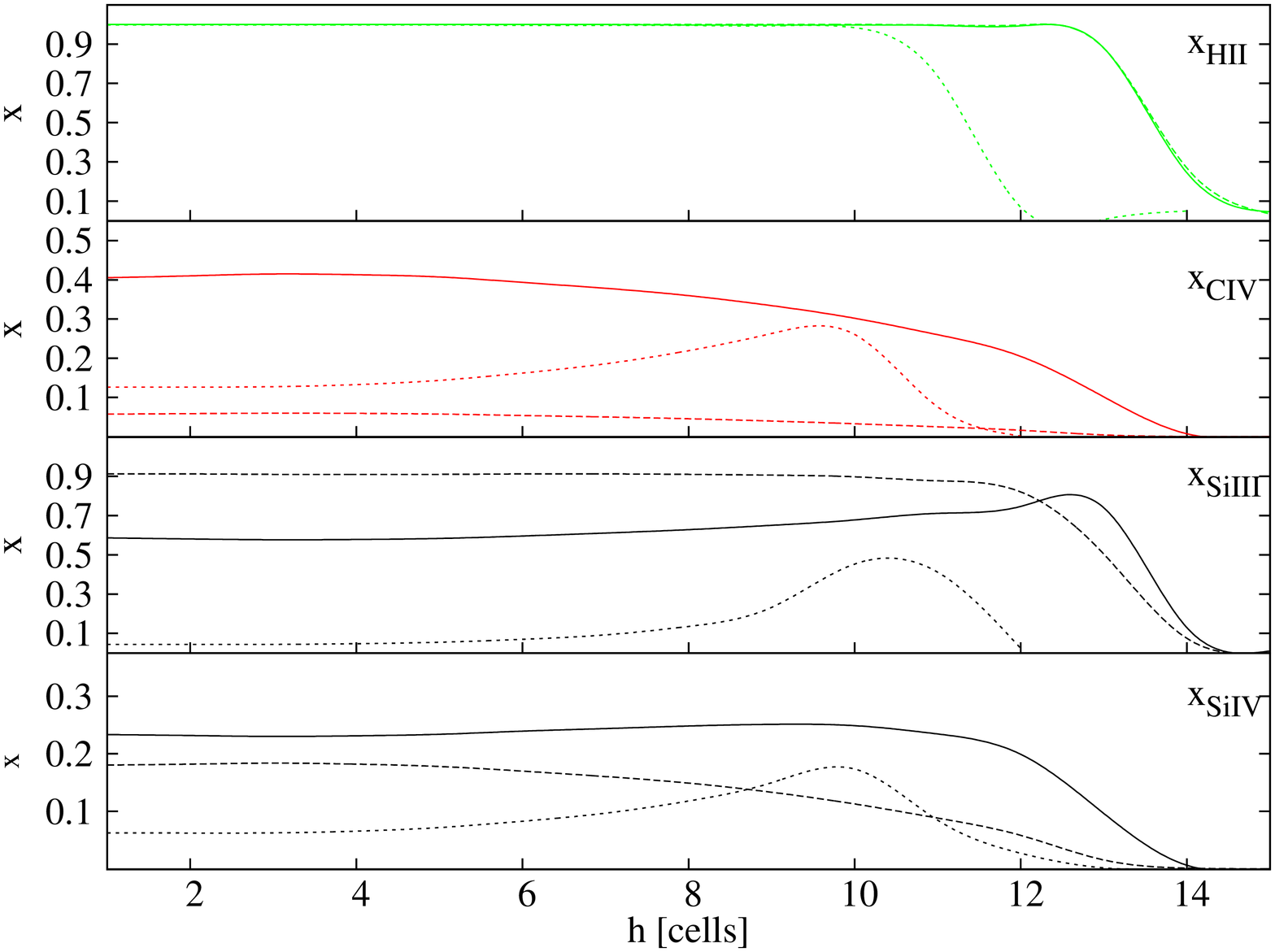}
\vspace{1.0truecm}
\caption{Ionisation fractions as function of $h$, evaluated in the plane equidistant
  from the two sources ($x=30$) of Test~2 when the source spectrum is          
  changed compared to the reference case (see text for details).
  From the top to the bottom the panels refer to $x_{\textrm{HII}}$,
  $x_{\textrm{CIV}}$, $x_{\textrm{SiIII}}$ and $x_{\textrm{SiIV}}$.
  In all panels solid lines refer to the reference case with $T_1=T_2=10^{5}$~K; the dashed to 
  $T_1=T_2=5\cdot10^{4}$~K and the short-dashed to $T_1=T_2=5\cdot10^{5}$~K.}
\label{fig:Behaviour-L1fix-L2fixTvar}
\end{figure}

\subsubsection{Variations in the source spectra}

In this Section we study the variations induced 
by changes in the temperatures $T_{1}$ and $T_{2}$ of
the black-body spectra. The source ionisation rates are set to the
reference value $\dot{N}_{1}=\dot{N}_{2}=9\cdot10^{51}\textrm{phot s}^{-1}$. 
The results in terms of distribution of some reference ionisation fractions
are shown in Figure~\ref{fig:Behaviour-L1fix-L2fixTvar}.

In the top panel the H$\, \rm \scriptstyle II$
I-front is identical in the cases $T_1=T_2=5\cdot10^{4}$~K and $T_1=T_2=10^{5}$~K,
while it recedes significantly for $T_1=T_2=5\cdot10^{5}$~K.
This is because by increasing the temperature from $5\cdot10^{4}$~K to $10^{5}$~K,
most of the photons still have $E<E_{\rm HeII}$ and get preferentially absorbed by
hydrogen because of its higher abundance. 
For these two cases $x_{\rm HeII}$ (which
is not plotted here) has a profile very similar to that of $x_{\rm HII}$, with the only 
difference that it slightly increases with increasing temperature because more photons
have $E>E_{\rm HeI}$. On the other hand, when the temperature of the black-body spectra
is increased to $5 \cdot 10^5$~K, most of the photons have an energy in the vicinity of $E_{\rm HeII}$,
causing a drop in $x_{\rm HII}$ and $x_{\rm HeII}$ and the appearance of $x_{\rm HeIII}$
(which is not plotted here), which was missing in the other cases, being relegated only
in the immediate surroundings of the sources.

While for $T_1=T_2=5\cdot10^{4}$~K there is hardly any C$\, \rm \scriptstyle IV$
available because of the lack of enough photons with $E>E_{\rm CIII}$, as the spectrum temperature increases so does $x_{\textrm{CIV}}$. Increasing the black-body temperature
further, and thus shifting the peak of the black-body spectrum to higher
frequencies, brings a drop in $x_{\textrm{CIV}}$, which is balanced by an equal
increment of $x_{\textrm{CV}}$ (which is not shown in this plot).

Unlike in the previous tests, in this case both Si$\, \rm \scriptstyle III$ and
Si$\, \rm \scriptstyle IV$
show remarkable changes by varying the spectrum temperature.
While $x_{\rm SiIII}$ keeps decreasing as the temperature increases and 
higher ionisation states are favoured, $x_{\rm SiIV}$ increases from
$T_1=T_2=5\cdot10^{4}$~K to $T_1=T_2=10^{5}$~K, while it decreases when
$T_1=T_2=5\cdot10^{5}$~K, similarly to $x_{\rm CIV}$.

Also in this case we can conclude that {\tt CRASH3} is very sensitive to changes in
the spectrum of the ionising sources.

 We have verified that a similar conclusion
applies when only one of the two source temperature is changed or when we have
adopted a power-law rather than a black-body spectrum. 

\subsection{Test 3: radiative transfer on a cosmological density field enriched
by metals \label{sub:TEST3} }

\begin{figure}
\centering
\includegraphics[width=0.50\textwidth]{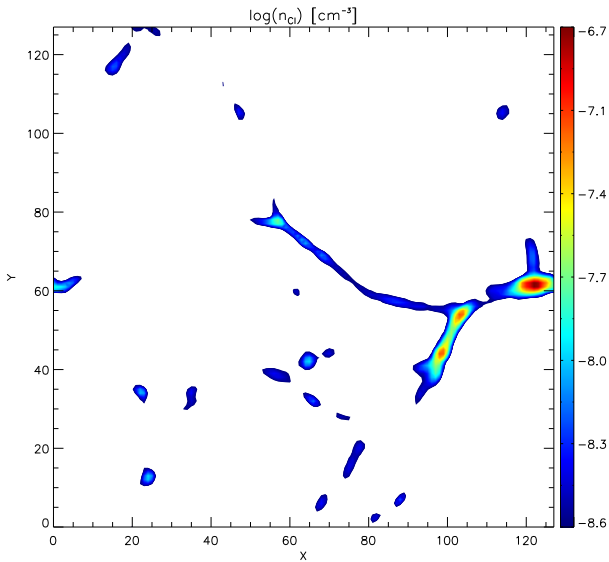}     
\caption{Slice cut through the simulation box showing the initial
  distribution of $\log(n_{\textrm{CI}})\,[\textrm{cm}^{-3}]$ in Test~3.
}
\label{fig:-Test3CITMaps1}
\end{figure}

In this Section we present an extension of the Test~4 proposed in the
Cosmological Radiative Transfer Comparison Project \citep{b18-Iliev2006a}.
The original test set-up has been adopted, but we have extended the 
ICs to include He, C, O and Si.
For reference, the box size is $L_{b}=0.5h^{-1}\textrm{Mpc}$ co-moving,
$h=0.72$, $N_{c}=128$, the simulation duration is set to $t_{f}=4\cdot10^{5}$~yrs,
starting at redshift $z=9$. 
The cosmological evolution of the box is disabled and the simulation starts with a neutral gas at initial temperature $T=100$~K.

The H and He fractions are the same as in Test 1.
16 point sources are present in the box with a black-body spectrum
at  $T=10^{5}$~K. The sampling of the spectrum is done with 91 frequencies,
to reach the accuracy established in our Test 1. Finally, we have
used $10^{8}$ packets to ensure a good convergence as done in MCK09. 
It is important to note that the aim of this test is to show how the
\texttt{CRASH3} pipeline is applied to a realistic density configuration
and the rich set of information provided by it, and it is not meant to quantitatively reproduce observational results.

Differently from the previous tests in which we populated the full
computational volume with metals, here we assign a metallicity of
$Z_{g}=\Delta \cdot 0.0064 Z_{\odot}$ to those cells with an overdensity
$\Delta   =\rho/\langle \rho\rangle>10$.
This results in about 5 percent of metal enriched cells. The number
density of C, O and Si relative to the hydrogen are the same          
ones introduced at the beginning of the Test Section. 

In Figure \ref{fig:-Test3CITMaps1} the distribution
of the C$\,{\rm {\scriptstyle I}}$ number density is shown in a slice of
the simulation box containing the brightest source, located within the densest
region in the computational volume. The distribution of Si$\,{\rm {\scriptstyle I}}$
and O$\,{\rm {\scriptstyle I}}$ is the same, with $n_{{\rm SiI}}=0.142\cdot n_{{\rm CI}}$
and $n_{{\rm OI}}=2\cdot n_{{\rm CI}}$. 
By construction, most of the metals are concentrated in the vicinity
of the sources and in general in the highest density regions.

As shown in Figure \ref{fig:TCorrectionMetallicity} of Test 1,
gas with metallicity  $Z_g > 0.1 Z_{\odot}$ starts to cool with different efficiency 
depending on the distance from the illuminating source (see comments in the Section), while 
for $Z_g > 0.64 Z_{\odot}$ the cooling is relevant at every distance from the source. 
In the box of this Test, cells with $\Delta>100$, are only 0.012 percent of the total number, 
while cells with $\Delta>50$ are about 4 percent. We then expect the effect of metal cooling to be 
negligible on the global ionisation status of the medium and we ignore it in discussing the results.
In realistic 3D simulations the metal cooling should be carefully accounted for in all the cells with 
$Z_g > 0.1 Z_{\odot}$, because it acts differently depending on the gas metallicity and the RT effects shaping the field. 
The statistical effects of the metal cooling will be analysed with care in future  
applications in the context of physically motivated enrichment patterns.

\begin{figure}
\centering
\includegraphics[width=0.50\textwidth]{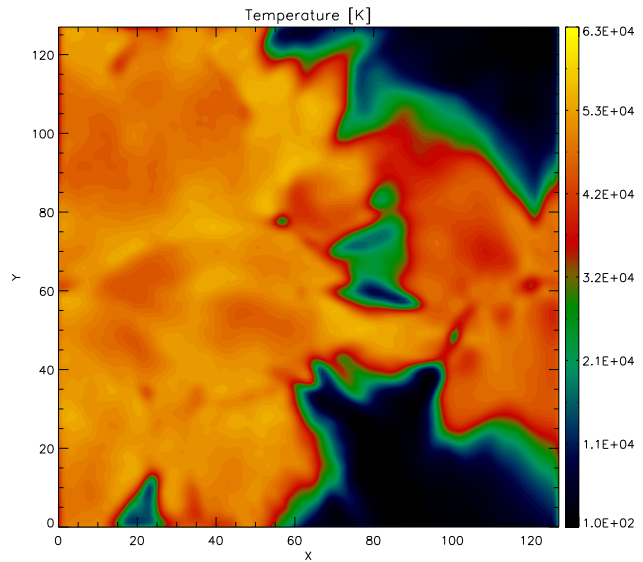}
\caption{Slice cut through the simulation box showing the
  distribution of the gas temperature $T$ in the same slice
  of Figure~\ref{fig:-Test3CITMaps1} at the time $t=5 \cdot 10^4$~yrs in Test~3.}\label{fig:-Test3CITMaps2}\end{figure}

\begin{figure*}
\centering
\includegraphics[width=0.8\textwidth]{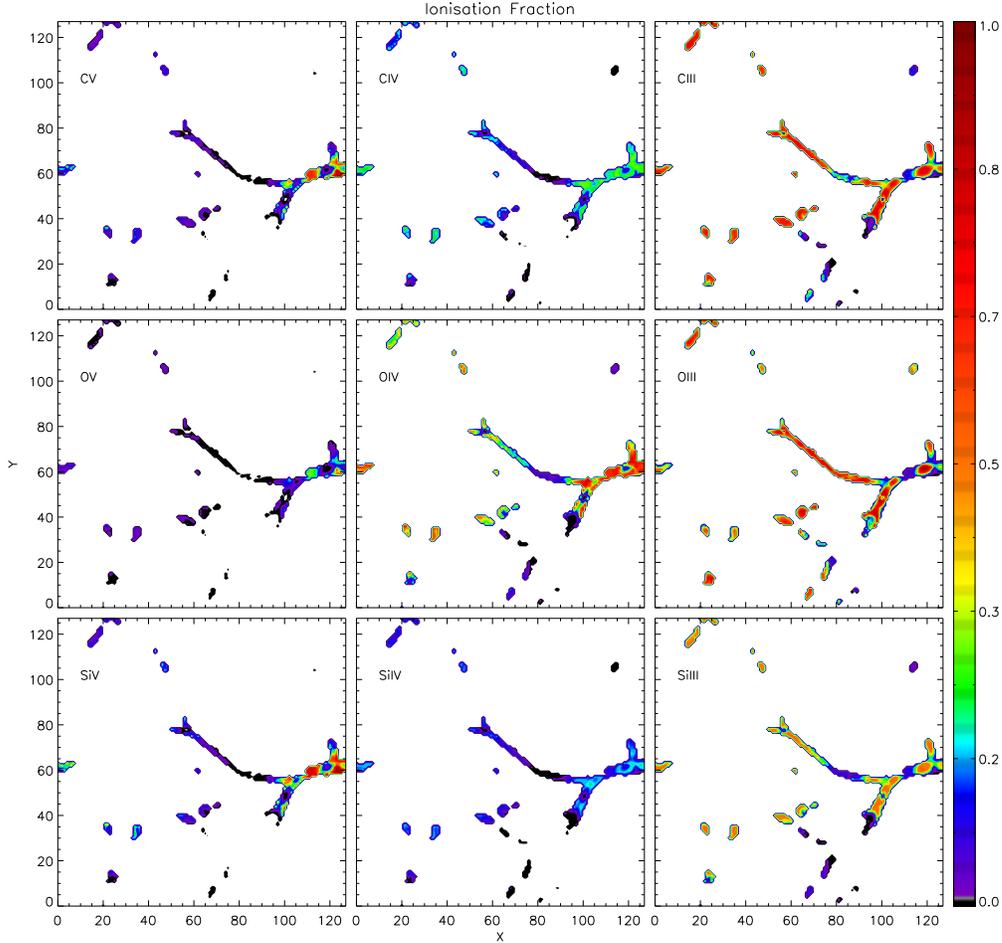}
\caption{Slice cut through the simulation box showing the distribution
  of (from left to right and from top to bottom): $x_{\rm CV}$, $x_{\rm CIV}$,
  $x_{\rm CIII}$, $x_{\rm OV}$, $x_{\rm OIV}$, $x_{\rm OIII}$,
  $x_{\rm SiV}$, $x_{\rm SiIV}$ and $x_{\rm SiIII}$. The slice shown is
  the same of the previous Figures at the time $t=5 \cdot 10^4$~yrs in Test~3.}
\label{fig:ion_metals}
\end{figure*}

To better understand the distribution of the metal ionisation species
discussed in the following, in 
Figure \ref{fig:-Test3CITMaps2} we show the distribution of gas temperature in
the same slice as above at a time $t=5 \cdot 10^4$~yrs.
Black regions ($T \sim 100$~K)
correspond to areas not reached by the radiation field, the green
color clearly traces the He$\, \rm \scriptstyle II$ ionisation
fronts at a typical temperature $T \sim 2 \cdot 10^{4}$~K, while
the red/orange is associated to regions at $T\sim 4 \cdot 10^{4}$~K,
dominated by H$\, \rm \scriptstyle II$ and He$\, \rm \scriptstyle II$.
Finally, warmer, yellow regions correspond to areas where there is also
substantial He$\, \rm \scriptstyle III$.

The distribution of some reference metal species is shown in Figure~\ref{fig:ion_metals} across 
the same slice.
While in the vicinity of the brightest source C$\, \rm \scriptstyle V$ is present in large
quantities, with $x_{\rm CV}$ as high as 1, only small traces are visible further away.
On the other hand, these regions are typically rich in C$\, \rm \scriptstyle III$, showing
a qualitative behaviour similar to the one discussed in Test~1, where C$\, \rm \scriptstyle V$
and C$\, \rm \scriptstyle III$ are complementary species. Consistently with the same Test~1, 
C$\, \rm \scriptstyle IV$ is present everywhere along the filaments with relatively low abundances. Regions
with fractions close to zero are those which have not been reached by ionising radiation
(see Fig.~\ref{fig:-Test3CITMaps2}).
A similar behaviour is observed for O, with O$\, \rm \scriptstyle V$ present in very small traces
only in the vicinity of the brightest source, O$\, \rm \scriptstyle IV$ extending a bit further away
along the filaments and O$\, \rm \scriptstyle III$ being present in substantial amounts elsewhere.
Finally, while there are only traces of Si$\, \rm \scriptstyle IV$, $x_{\rm SiV}$ is as high as 1 in the
vicinity of the source and Si$\, \rm \scriptstyle III$ is present in modest quantities everywhere.

While it is not possible to make a quantitative comparison with Test~1, the qualitative results
in terms of ionisation fractions at a given distance from the brightest source are in good agreement,
with the highest ionisation levels present only in the close proximity of the source and the lower
ionisation levels present everywhere along the filaments hit by the ionising radiation.

\begin{figure*}
\centering
\includegraphics[width=0.8\textwidth]{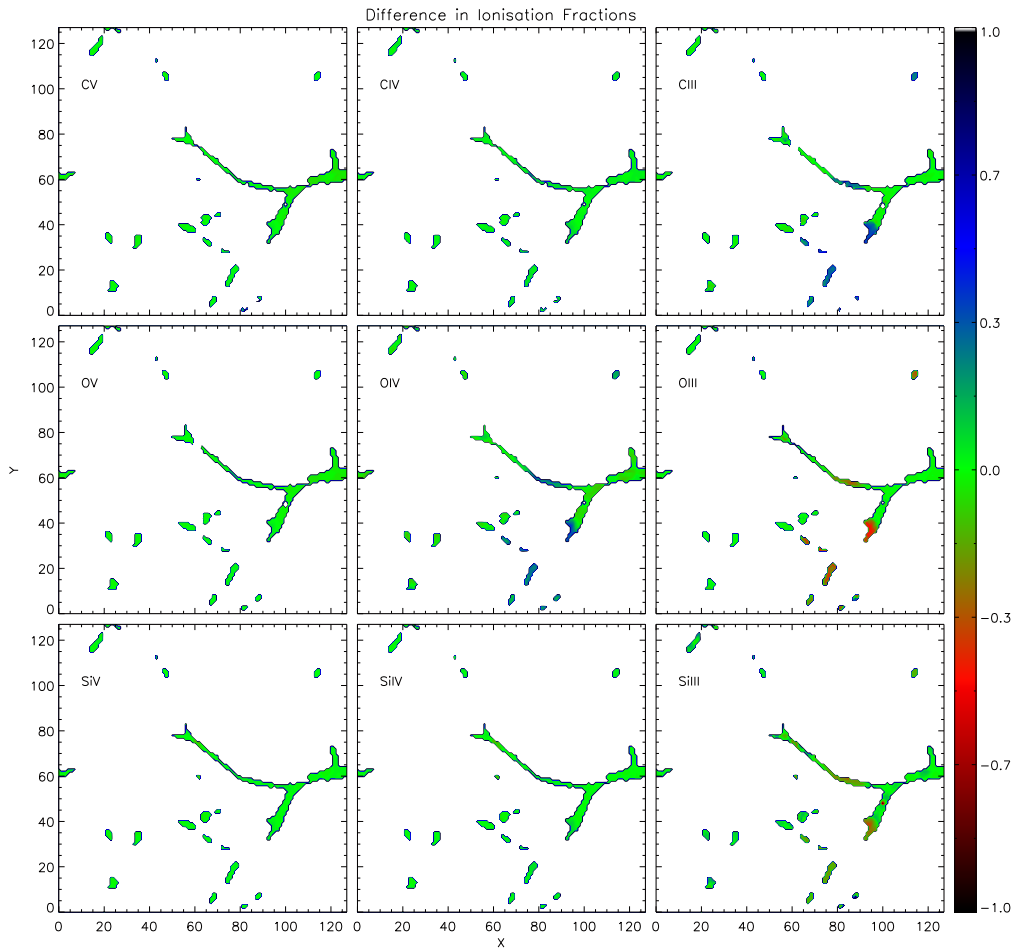}
\caption{Slice cut through the simulation box showing the distribution
  of the differences in ionisation fractions obtained in Test~3 with the reference
  source spectrum and with a spectrum extending to a lower ionising energy of $E_{\rm SiI}$
  ($\Delta x_i = x_{i, E_{\rm SiI}} - x_{i, E_{\rm HI}}$; see text for more details). 
  From left to right and from top to bottom the panels refer to the species: C$\, \rm \scriptstyle V$, 
  C$\, \rm \scriptstyle IV$, C$\, \rm \scriptstyle III$, O$\, \rm \scriptstyle V$, 
  O$\, \rm \scriptstyle IV$, O$\, \rm \scriptstyle III$, Si$\, \rm \scriptstyle V$, 
  Si$\, \rm \scriptstyle IV$ and Si$\, \rm \scriptstyle III$. The slice shown is
  the same of the previous Figures at the time $t=5 \cdot 10^4$~yrs.}
\label{fig:ion_metals_difference}
\end{figure*}

Similarly to what done in Section~\ref{sec:lowerenergy}, we have investigated the impact on the
evaluation of metal ionisation states of photons with $E<13.6$~eV by extending the spectrum to
to energies lower that $13.6$~eV down to $E_{\rm SiI}$. 
In Figure~\ref{fig:ion_metals_difference} the differences in ionisation fractions 
($\Delta x_i = x_{i,E_{\rm SiI}}-x_{i,E_{\rm HI}}$) computed in the two set-ups are reported 
for the same ions of  Figure~\ref{fig:ion_metals}.  
As for Test~1, we find that where $x_{\rm HII} \sim 1$ the agreement
between the case with and without the inclusion of the low energy tail is excellent in 
most of the metal enriched domain.
$x_{\rm OIV}$ and $x_{\rm OIII}$ present an average discrepancy of about 13\%, with a maximum
of $\sim$ 40\% confined in few cells where the ionisation fractions are relatively low 
(i.e. below $\sim$ 30\%). For higher $x$ the agreement is excellent, reflecting the results
in Section~\ref{sec:lowerenergy}, where the largest discrepancies were observed in correspondence 
of those cells in which the abundance of the different species was raising or declining. 
On the other hand $x_{\rm OV}$ always shows a good agreement between the two cases. 
The Si and C ionisation fractions present a maximum discrepancy below 10\% for 
their higher (i.e. $\, \rm \scriptstyle IV$ and $\, \rm \scriptstyle V$) ionisation species, 
while it raises to $\sim$40\% for
$x_{\rm CIII}$ and $x_{\rm SiIII}$. Also for these species only few cells have
such large discrepancies. 

In general we can conclude that, similarly to what seen in Test~1, the extension of the spectrum to energies
lower than 13.6~eV results in some discrepancies in the evaluation of the metal ionisation fractions.
Such discrepancies are limited to some of the metal ions (especially the lowest ionisation states) and are largely due to a number of physical processes which are neglected (see footnote of Test 1 and the discussion in Section~\ref{sec:lowerenergy}). In the selected slice, they are also confined to a small number of cells typically located far from the ionising sources. As a reference, the percentage of metal enriched cells
which present a discrepancy larger than 5\%, 10\%, 25\% and 50\% in the visualised species is 19.3\%, 9.6\%, 
4.2\% and 0.6\%, respectively.

\section{CONCLUSIONS}

In this paper we presented \texttt{CRASH3}, the latest release of the 3D radiative transfer code \texttt{CRASH}.
In its current implementation \texttt{CRASH3} integrates into the reference algorithm the code \texttt{Cloudy} to evaluate the ionisation states of metals, self-consistently with the radiative transfer through the most
abundant species, i.e. H and He.  The feedback of the heavy elements on the calculation of the gas temperature is also taken into account, making of {\tt CRASH3} the first 3D code for
cosmological applications which treats self-consistently the  radiative transfer through an
inhomogeneous distribution of metal enriched gas with an arbitrary number of point sources and/or
a background radiation.
It should be noted that the algorithm is based on the assumption that metals do not 
contribute to the gas optical depth, which is valid in typical configurations of cosmological interest, as
e.g. in the IGM. On the other hand, the code is not suitable for applications in which metallicities \gsim $Z_\odot$ are involved.

The pipeline has been tested in idealized configurations, such as the classic Str\"omgren sphere, albeit
enriched with heavy elements, as well as in a more realistic case of multiple sources
embedded in a polluted cosmic web, which shows the rich set of information provided by the
metal ionisation states.
Through these validation tests the new method has been proved to be numerically stable and 
convergent with respect to a number of parameters.

The dependence of the results on e.g.  the source characteristics (spectral range
and shape, intensity), the metal composition, the gas number density and metallicity has been investigated.
Despite the difficulty in testing quantitatively the results, we find that the qualitative behaviour is consistent with our expectations.

In conclusion, {\tt CRASH3} is an excellent code to simulate the evolution of various species in a
low metallicity gas illuminated by  an ionising radiation, which can be modelled both as a background
radiation and point sources. {\tt CRASH3} will then be an invaluable
tool to investigate the interpretation of e.g. metal absorption lines in quasars' spectra or 
fluctuations in the UVB, in more detail, with a better modelling of the relevant physical processes and with a 
higher accuracy than done before. These applications are topics for further investigations.

\section*{ACKNOWLEDGMENTS}
The authors would like to thank the referee, Alexey Razoumov, for his very constructive
comments. The authors are also grateful to J. Bolton, R. Dav\'{e}, A. Ferrara and J. Ostriker
for enlightening discussions. AM acknowledges the support of the DFG
Priority Program 1177. LG acknowledges the support of the DFG Priority Program 1573.

\begin{appendix}

\section{Convergence tests}
When the temperature feedback is introduced in the {\tt CRASH3} pipeline 
(see description of Step 3) $k_f$ times, the RT process is altered. While the gas recombination depends weakly on 
the temperature and we can expect negligible effects, the numerical convergence of the metal ionisation fractions 
needs be verified for different values of $k_f$. 
Here we show the results of some tests run to check the convergence of our results with respect
to $k_f$ when the metal feedback is enabled.

\begin{figure}
\centering
\includegraphics[angle=-90,width=0.50\textwidth]{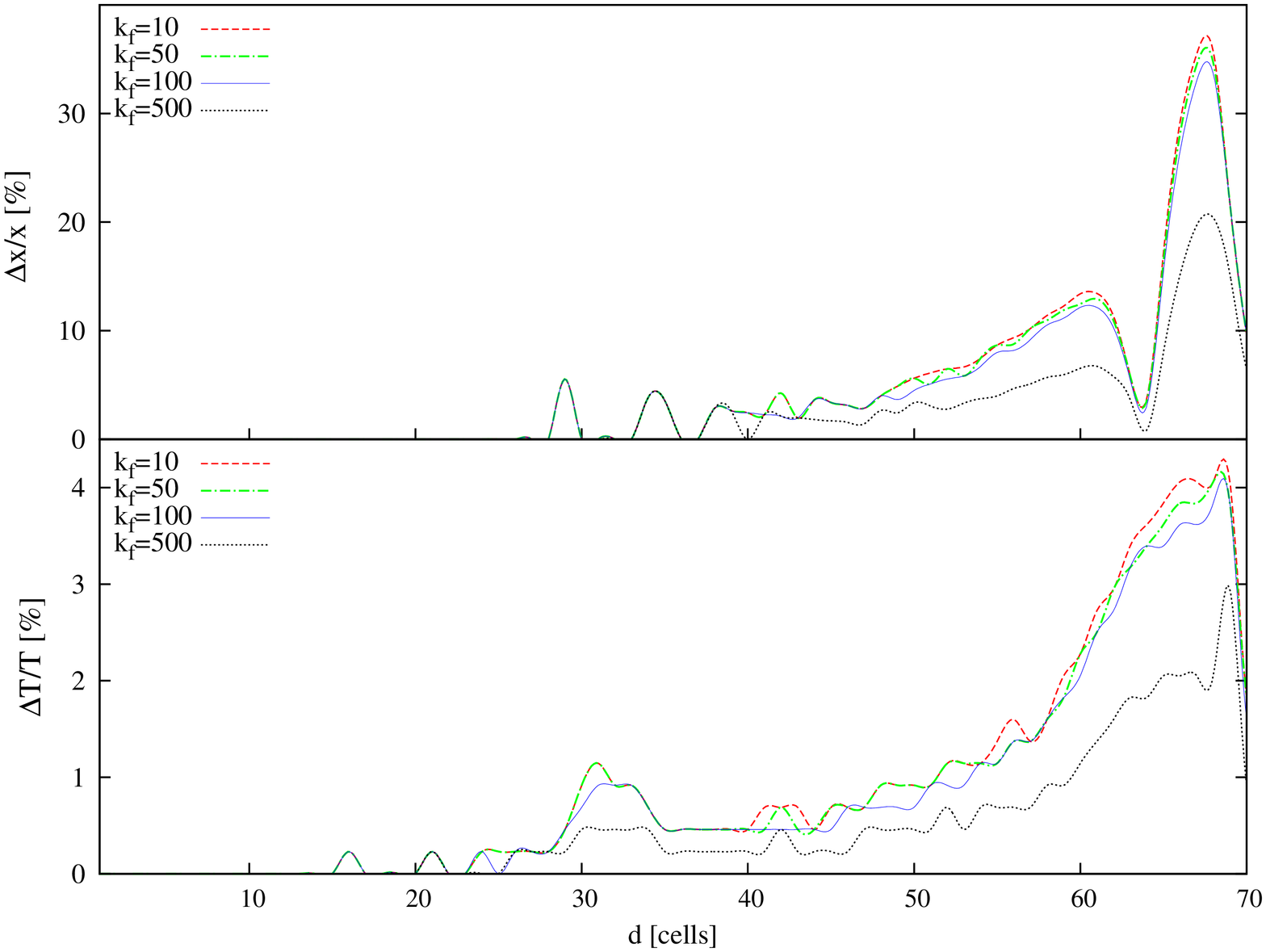}
\vspace{0.5truecm}
\caption{$\Delta x/x$=$\mid$$x(k_f$=$1000)$-$x(k_f)$$\mid$/$x(k_f$=$1000)$ ({\bf upper panel})
  and $\Delta T/T$=$\mid$$T(k_f$=$1000)$-$T(k_f)$$\mid$/$T(k_f$=$1000)$ ({\bf lower}) as a function of
  the distance $d$ from the source in Test~1 including the feedback from metals. Here $x$ refers to 
  the ionisation fraction $x_{\rm CII}$. The
  curves are plotted at the time $t_f=5 \cdot 10^8$~yrs and refer to a value of $k_f=10$
  (red dashed lines), 50 (green dashed-dotted), 100 (blue solid) and 500 (black dotted).}
\label{fig:FeedbackConvergence}           
\end{figure}

In Figure~\ref{fig:FeedbackConvergence} we show the profile of the variation of a reference ionisation fraction ($x_{\rm CII}$) and temperature for Test~1 run with the metal feedback switched on.
The lines refer to the final time $t_f=5 \cdot 10^8$~yrs and to values of $k_f=10$, 50,
100 and 500. 
The convergence of the results within the ionised sphere ($d \leq 40$ cells) is excellent: even for small values of $k_f$ the relative discrepancies remain within few percent both in the ionisation fraction (top panel) and in the temperature (bottom panel).
Numerical oscillations of a few percent (or smaller) in $\Delta x/x$ are present in 
a handful of isolated cells, but they disappear for $k_f \geq 100$. 

Closer to the I-front ($d \geq 50$), where $x_{\rm CII}$ is very small (i.e. $\leq 10^{-4}$),
the convergence is worse, in particular in the 2-3 cells encompassing the I-front. This behaviour was expected because, as already discussed in Test 1, the accuracy of our scheme is diminished in the outer
regions of the ionised bubble, where the \texttt{Cloudy} computation is not always convergent. 
It should be noticed, however, that $x_{\rm CII}$ presents one of the worse behaviours and higher 
ionisation states are more numerically stable.  
We have checked that the convergence is similar also at earlier times, when the ionised sphere has not reached an equilibrium configuration.
Other species show a  similar dependence on $k_f$. For example, $x_{\rm CIV}$ presents an excellent convergence (of the order of a percent) within the ionised region already with $k_f=10$, while in the proximity of the I-front, where  $x_{\rm CIV} \leq 10^{-6}$, it degrades to tens of percent. 

The temperature shows an even better behaviour, with a relative convergence of a few percent even in the
vicinity of the I-front. Also at earlier times the convergence is never worse than 10\%. The oscillations in the
curves have the same origin of those in the upper panel.

We should notice that the value of $k_f$ needed to reach a good convergence is likely dependent on the
problem at hand. In general, though, we can conclude that we expect a value of $k_f \sim 500$ to be
sufficient for most applications which require the feedback of metals on the determination
of the gas temperature.

\section{Database implementation and lookup strategy}

Here we provide more details on the implementation of the relational 
database introduced in Section~4 
and the lookup strategy we use to query its data.

The introduction of a relational database storing precomputed data is motivated by the huge number 
of \texttt{Cloudy} computations ($N_{comp}$) needed by our pipeline in specific metal 
configurations. Consider a general reionisation simulation involving $N_{z}$ 
density snapshots in which the metals are computed $t_{f}$ times on
a sub-domain of $N_{m}(z)$ cells; the total number of computations ($N_{comp}$) is:

\begin{equation}
N_{comp} = N_{z} \times N_{m}(z) \times t_{f}.\label{eq:Num_Comp}
\end{equation}

Note that $N_{comp}$ depends indirectly on the adopted grid resolution $N_{c}$ and the metal filling 
factor, therefore the size of the metal map $N_{m}$ can rapidly reach a high value.
For instance, a simulation with $N_{c}=128^3$, $N_{m} \sim 10^5$ (i.e. cell filling 
factor of 5\%) and $t_{f}=10$ requires $N_{comp} \sim 10^6$ computations in a single snapshot. 
As the direct evaluation of $N_{comp}$ \texttt{Cloudy} runs during the RT simulation 
is feasible just in few cases, we use a DB of precomputed configurations to alleviate 
the total computational cost. 
    
The DB stores a large number of \texttt{Cloudy} runs abstracted as object-data associating 
initial conditions (RUN\_ICs) with the resulting ionisation fraction and temperature  (RUN\_RESULTS); 
this association is uniquely identified by an indexing key (RUN\_ID).

The set of RUN\_ICs is composed by the number densities of all the species $n_{m}$ polluting the metal 
contaminated sub-domain, the luminosity $L_m$ 
and the spectral shape $S_m$, specified on a fixed number of frequency bins (see  Section 4). 
As detailed in the Test Section, 91 bins are generally required to obtain the 
accuracy of the tests in matching the HII fronts.

The RUN\_RESULTS set is composed by the ionisation fractions of all the species $x^{C3}_m$
and the temperature $T^{C3}$.

As prescribed by the relation database theory, all the variables are organised in columns 
of the database tables and related by the RUN\_ID. The database solution is practically 
implemented by combining two relational databases: a reduced "in-memory DB" (called production DB)
used during the RT (see Step 3 of the Pipeline in Section 4.3), and an "archiving DB" 
(a standard TCP/IP accessible DB) collecting the results of many runs and increasing in time. 
Before running a simulation the content of the smaller DB is extracted from the archiving DB 
and customised on the problem at hand as explained below.
   
The content of the production DB can be tuned on the specific problem under investigation because the \texttt{CRASH3} 
pipeline is applied in post-processing and part of the RUN\_ICs can be partially pre-constrained. 
First, the $n_{m}$ values are exactly known from the hydro-simulation at fixed redshift, while the values 
of $L_m$ and $S_m$ can be only predicted because they are affected by the metal feedback.
Second, in the regime of applicability of our pipeline we tested that the metal cooling feedback is weak on the 
RT field and by running a RT simulation just through H and He (the only absorbers, see Section 4) we can 
have a reasonable estimate of the statistics of the produced spectral shapes. 
Note also that in absence of feedback from metals $L_m$ and $S_m$ are provided exactly from Step 2 at every time
but $N_{comp}$ can be huge and still require some pre-computed configurations to successfully finish the run.

The DB is then populated by spanning a subset of compatible ICs providing a non-redundant set of 
final ionisation fractions and temperatures; this is realised by testing the sensitivity of the 
photo-ionisation software to small variations ($\delta n$, $\delta L$, $\delta S$) of their values. 
An example of these tests is reported in Section 5.1 when we studied the effects of the metal 
feedback on the resulting temperature and ionisation fractions (see Figures 9 and 10). 

During a CRASH3 simulation, the pipeline is responsible for lookup the DB at Step 3 as explained in 
Section 4.3. 
To recognise a compatible pre-computed configuration in each $m$-cell of the metal polluted sub-domain, 
Step 3 performs a cascade of SQL-SELECT queries by comparing the RUN\_ICs of the interested cell with 
the RUN\_ICs records stored in the DB tables, as explained below.
      
The first query (Q\_1) selects a ResultSet (R\_1) by querying for the compatible number 
densities: $n_m \pm \delta n_m$. The number of records (i.e. table rows) in R\_1 is defined as $R_1$.
This is realised by a series of \texttt{SQL-BETWEEN} conditions on $n_m$, 
concatenated by \texttt{.AND.} clauses; in SQL meta-code:  

\begin{lstlisting}

SELECT FROM < TABLE_NAME > 

  WHERE (

   (nHI BETWEEN nHI + - dnHI) 

  AND 

   (nHeI BETWEEN nHeI + - dnHeI) 
  
  AND ...

  )

\end{lstlisting}

A second query Q\_2 is performed on R\_1 applying the same logic to the cell luminosity $L_m$ and providing a 
ResultSet R\_2 ($R_2 \ll R_1$) in which $L_m \in L_m \pm \delta L_m$.

Finally, a Q\_3 query finds compatible spectral shapes $S_m$ in the 
ResulSet R\_2 by simultaneously comparing the shape in every spectral bin within the tuned 
oscillation $\delta S$. Q\_3 provides a ResultSet R\_3 with $R_3 \ll R_2$.
This last query involves all the spectral bins and drastically reduces the number $R_3$ of final candidate configurations. 
On the other hand it is necessary to strictly compare the spectral shapes in every bin to guarantee that the spectral fluctuations due to RT effects are correctly accounted for and the full compatibility between results coming from the DB queries and the ionisation fractions produced by the complementary on-the-fly runs.
 
The set R\_3 of candidate pre-computations is further reduced to $R_3=2$ final candidates by 
minimising the difference of their  H and He ionisation fractions with the values found at 
Step 2 (See Section 4.3). We generally allow a maximum difference below 20 \% for H and
10 \% for both He fractions, but these values can be adapted to the problem at hand, increasing 
the precision of the matching criterion.
The final result is then found as mid-point of the linear interpolation of their H,He and metal 
fractions if they are either sides of the Step 2 results, otherwise just the configuration with the closer value is considered, 
first checking for $x_{\textrm{HeII}}$ and $x_{\textrm{HeIII}}$, and finally for $x_{\textrm{HII}}$. 
As an example of the accuracy obtained in the resulting configuration consider the convergence test in Section 5.1.1 
for a HII region. 

Also notice that in all the cells with $x_{\textrm{HII}}=x_{\textrm{HeII}}=x_{\textrm{HeIII}}=1$ (e.g. the first 8 cells of Figure 2),
the configurations in the set R\_3 cannot be disentangled. In this case either the metal ionisation fractions differ within a threshold 
value (normally $10^{-3}$) and one of them is accepted, or the entire R\_3 set is rejected and a on-the-fly run is performed.

As final comment we point out that the DB solution has the only purpose of reducing the computational complexity of the problem
by reusing previous computations and cannot provide any general solution of the RT through metals, but must be targeted on the 
specific problem at hand. 
Also note that for a maximum accuracy in the metal ion predictions, and when the value of $N_{comp}$ implies a reasonable computational time, 
the DB matching thresholds can be reduced or the DB lookup can be excluded from the pipeline allowing just on-the-fly computations. 
This is the case of most single snapshot computations with medium resolution and low metal filling factor.       

\end{appendix}

\label{lastpage}
\end{document}